\documentclass[prb,twocolumn,floatfix,epsf,amssymb,longbibliography,aps]{revtex4-1}
\usepackage{graphicx,subfigure,amsmath,colordvi,times}
\usepackage[usenames]{color}
\usepackage[english]{babel}
\usepackage{amstext}
\usepackage{amsfonts}
\usepackage{amssymb}
\usepackage{bm}
\usepackage{float}
\usepackage{xcolor} 
\usepackage{hyperref}
\hypersetup{pdfborder=0 0 0,colorlinks=true,citecolor=blue,linkcolor=blue}

\begin{document}

\date{\today}
\title{Experimental conditions for observation of electron-hole superfluidity in GaAs heterostructures}
\author{Samira Saberi-Pouya,$^1$ Sara Conti,$^{1,2}$ Andrea Perali,$^3$ Andrew F. Croxall,$^4$ Alexander R. Hamilton,$^5$
Fran\c{c}ois M. Peeters,$^{1}$ and David Neilson$^{1,5}$}
\affiliation{$^1$Department of Physics, University of Antwerp, Groenenborgerlaan 171, 2020 Antwerpen, Belgium\\
${^2}$Physics Division, School of Science and Technology, Universit\`a di Camerino, 62032 Camerino (MC), Italy\\
${^3}$Supernano Laboratory, School of Pharmacy, Universit\`a di Camerino, 62032 Camerino (MC), Italy\\
${^4}$Cavendish Laboratory, University of Cambridge, J.J. Thomson Avenue, Cambridge CB3 0HE, United Kingdom\\
${^5}$ ARC Centre of Excellence for Future Low Energy Electronics Technologies, School of Physics,
The University of New South Wales, Sydney, New South Wales 2052, Australia}

\begin{abstract}
	The experimental parameter ranges needed to generate superfluidity in  optical and drag experiments in GaAs double quantum wells are
	determined, using a formalism that includes self-consistent screening of the Coulomb  pairing interaction in the presence of the superfluid. 
	The very different electron and hole masses in GaAs make this a particularly interesting system for superfluidity, 
	with exotic superfluid phases predicted in the BCS-BEC crossover regime. 
	We find that the density and temperature ranges for superfluidity cover the range for which optical experiments have observed 
	indications of  superfluidity, but that existing drag experiments lie outside the superfluid range.  
	We also show that for samples with low mobility 
	with no macroscopically connected superfluidity, if the superfluidity survives in randomly distributed localized pockets, 
	standard quantum capacitance  measurements could detect these pockets.  
\end{abstract}
\maketitle

While Bose Einstein Condensation (BEC) and the BCS-BEC crossover phenomena in superfluidity have been extensively studied  
for ultracold Fermi atoms,\cite{Bartenstein2004,Regal2004,Zwierlein2004}  it is probable that 
practical applications will instead be based on superfluidity in solid state devices.  
Existence of superfluidity in coupled  atomically-flat layers in semiconductor heterostructures
has been theoretically predicted,\cite{Perali2013,Conti2019b} while   
recent observations of dramatically enhanced tunneling at equal densities 
in electron-hole double bilayer sheets of graphene\cite{Burg2018,Efimkin2019} 
and in double monolayers of transition metal dichalcogenide monolayers\cite{Wang2019,Chaves2019}  
are strong experimental indications for electron-hole condensation.\cite{Spielman2000}  

Electron-hole superfluidity and the BCS-BEC crossover was first proposed for an excitonic system 
in a conventional semiconductor heterostructure of double quantum-wells in GaAs.\cite{Comte1982}
This was based on extensions of earlier work on exciton condensation.\cite{Keldysh1965,Keldysh1968,Lozovik1975,Lozovik1976}
To block electron-hole recombination, Refs.\ \citenum{Lozovik1975,Lozovik1976} 
proposed spatially separating the electrons and holes in a heterostructure consisting of two layers separated by an insulating barrier.

Superfluidity in GaAs quantum-wells differs in significant ways from superfluidity in coupled  atomically-flat layers.  
The large band gap in GaAs eliminates the multicondensate effects and multiband screening that are important in graphene,\cite{Conti2019} 
and the low-lying conduction and valence bands are nearly parabolic, and not dependent on gate potentials.    

However it is the widely different electron and hole effective masses that provides the most dramatic contrast of superfluidity in GaAs
compared with superfluidity in other solid state devices.  In GaAs the masses differ by a large factor:   
we take $m_{e}^\star=0.067m_{e}$ and $m_{h}^\star=0.3m_{e}$.  
Not only does this  have  significant consequences for the superfluid properties ,\cite{Saberi2018} but also for 
the screening responses of the electrons and holes, which are significantly different from the equal mass case.  
In ultracold atomic gases, Dy-K Fermi mixtures have been used to explore the physics of mass-imbalanced 
strongly interacting Fermi-Fermi mixtures.\cite{Ravensbergen2018}

The large mass difference makes double quantum wells in GaAs a solid state 
system uniquely suitable for generating and enhancing exotic superfluid phases that span the BCS-BEC crossover.\cite{Pieri2007}
Such phases can also be expected in mass-imbalanced ultracold atomic gas Fermi mixtures,\cite{Kinnunen2018} 
but only at currently inaccessible temperatures,\cite{Frank2018} $T_c \lesssim 50$ nK.  
The phases include the Fulde-Ferrell-Larkin-Ovchinnikov (FFLO) phase\cite{Wang2017a} 
and the Sarma phase with two Fermi surfaces (breached pair phase).\cite{Forbes2005}  
For GaAs, our estimates for  transition temperatures to the FFLO phase are readily accessible experimentally, $T_c \sim 0.2$ - $0.5$ K.
Potentially even more exciting is the possibility of a Larkin-Ovchinnikov supersolid phase 
when the masses are unequal.\cite{Baarsma2013,Zhang2019} 

For these reasons, experimental realization of superfluidity in GaAs quantum wells is of great interest.  
A major challenge facing experiments is that electron-hole superfluidity in double layer systems is exclusively 
a low density phenomenon, because strong screening of the long-range Coulomb pairing interactions 
suppresses superfluidity above an onset density $n_0$, 
and this is low.\cite{Perali2013,Neilson2014,LopezRios2018,Conti2019}  
Nevertheless, there are reports suggesting possible experimental signatures of electron-hole superfluid condensation 
in GaAs double quantum-wells.  Some signatures are based on optical observations of indirect exciton 
luminescence,\cite{Butov2002,High2012,Anankine2017,Timofeev2007} 
while others are based on transport measurements of Coulomb drag.\cite{Croxall2008,Seamons2009}

In this paper we map out the parameter space for GaAs double quantum well heterostructures 
to determine where electron-hole superfluidity is favored.   Important parameters are 
the widths $w$ of the quantum wells, 
the thickness $t_B$ of the Al$_x$Ga$_{1-x}$As insulating barrier,  
the densities in the wells $n$ (assumed equal), 
and any perpendicular electric fields.    
The distance between the peaks of the density distributions of the electrons and holes 
must be calculated, and is usually not simply the distance between the centers of the quantum wells.
By establishing the parameter ranges expected for superfluidity, we are able to provide independent corroborative support 
for the reported experimental signatures suggesting superfluidity.

Existing optical experiments generally use samples with quantum wells and barriers which are narrower than in  
samples for transport measurements.  A major reason is that  a problem arises for transport with the narrower wells  
from interface roughness scattering caused by 
Al atoms in the insulating barrier diffusing into the well regions.  Interface roughness scattering dramatically 
reduces the mobility, which is an essential consideration for transport measurements.   
Also, in optical measurements, electron-hole pairs are optically excited in a quantum well and then spatially separated 
across the barrier by means of a perpendicular electric field. 
Thus barriers are generally thinner than those for transport experiments, so the coupling of the electron-hole pairs
tends to be stronger for optical measurements.   

For the optical experiments we consider the samples from Refs.\ \citenum{Butov2002,High2012,Anankine2017}, with 
$8$ nm GaAs quantum wells separated by a $4$ nm barrier of Al$_x$Ga$_{1-x}$As, and   
Ref.\ \citenum{Timofeev2007} with $12$ nm wells and a $1.1$ nm AlAs barrier.  
Techniques to optically identify macroscopic spatial coherence were: the appearance in photoluminescence
measurements of bright localized spots with enhanced luminescence at fixed points on the sample;\cite{Butov2002}
the abrupt appearance of a sharp inter-well exciton line in the photoluminescence spectra;\cite{Timofeev2007}  
an abrupt increase in the amplitude of interference fringes using a Mach-Zehnder interferometer\cite{High2012} 
indicating a strong enhancement of the exciton coherence length; 
quenching of photoluminescence emission as a manifestation of optically-dark exciton condensation.\cite{Anankine2017}  
Indications of coherent condensation were observed at temperatures of a few Kelvin 
at carrier densities of a few $10^{10}$ cm$^{-2}$. 

\begin{figure}
	\centering
	\includegraphics[angle=0,width=1\columnwidth] {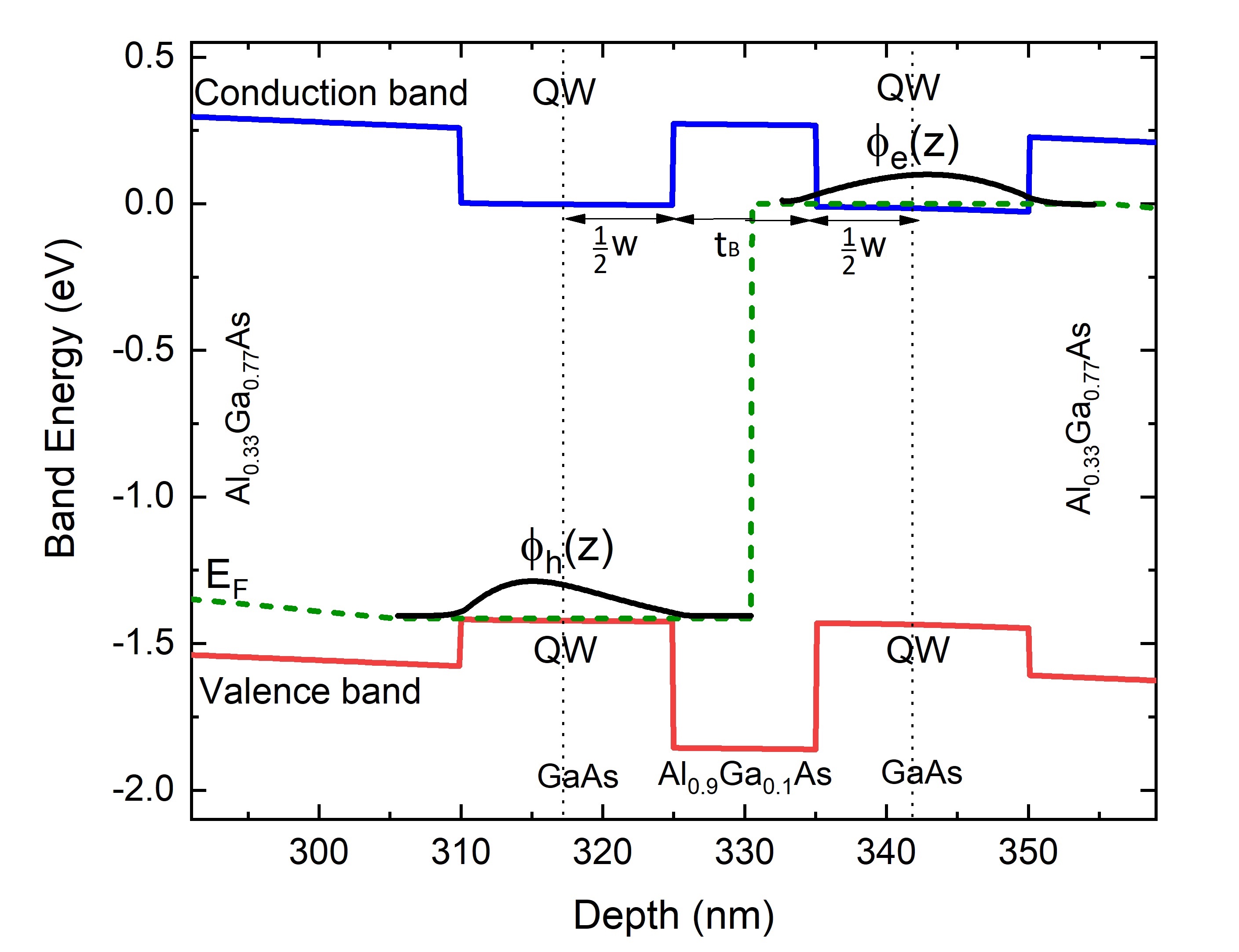}
	\caption{(Color online). Conduction and valence bands for a sample from Ref.\ \citenum{Croxall2008} 
		in the presence of gate potentials and a bias between the wells,\cite{Zheng2016}
		as obtained using a self-consistent Poisson-Schr\"{o}dinger solver:\cite{Tan1990} 
		quantum well widths: $w=15$ nm and Al$_{0.9}$Ga$_{0.1}$As barrier thickness: $t_B=10$ nm.  
		Dashed green line: Fermi level $E_F$. Vertical back dotted lines mark the centers of the wells.  
		$\phi_e(z)$ and $\phi_h(z)$ are the resulting electron and hole single-particle wave-functions confined in the wells.
		Note that the separation of the peaks in the $\phi_e(z)$ and $\phi_h(z)$ is larger than the distance between the centers of the two wells.
	}
	\label{Fig.1}
\end{figure}

From the samples used in the Coulomb drag experiments,\cite{Croxall2008,Seamons2009}
we examine the narrowest $15$ nm wells with the thinnest $10$ nm Al$_{0.9}$Ga$_{0.1}$As barrier (see Fig.\ \ref{Fig.1}).     
References \citenum{Croxall2008,Seamons2009} observed a  jump in the drag transresistivity around temperatures $T \sim 0.2$ - $1$ K.  
A sudden jump can be a signature of a superfluid transition,\cite{Vignale1996} but  
the observed deviations were not monotonic, sometimes even changing sign, so any signature of condensation was ambiguous. 

We start with the Hamiltonian,
\begin{eqnarray}
	\!\!\!\!\! {\cal{H}}\!=\!\! \sum_{\ell\mathbf{k}}\!\! \xi_{\mathbf{k}}^{\ell} 
	c^{\ell\dagger}_{\mathbf{k}} c_{\mathbf{k}}^{\ell}
	+\frac{1}{2}\!\!\!\sum_{\substack{\ell \neq  \ell',\\ \mathbf{q}, \mathbf{k},\mathbf{k}'}} 
	\!\!\!V^{\ell\ell'}_{\mathbf{k}-\mathbf{k}'}
	c^{\ell\dagger}_{\mathbf{k}+\frac{\mathbf{q}}{2}}
	c^{\ell'\dagger}_{-\mathbf{k}+\frac{\mathbf{q}}{2}}
	c^{\ell}_{-\mathbf{k}'+\frac{\mathbf{q}}{2}}
	c^{\ell'}_{\mathbf{k}'+\frac{\mathbf{q}}{2}},
	\label{Grand-canonical-Hamiltonian}
\end{eqnarray}
where $c^{\ell\dagger}_{\mathbf{k}}$ and $c_{\mathbf{k} }^{\ell}$ are 
creation and destruction operators with label $\ell=e\ (h)$ 
for electrons (holes) in their respective quantum wells, and   
$\xi_{\mathbf{k}}^{\ell}= {k}^{2}/(2 m_{\ell}^\star)- \mu^{\ell}$ are the 
single-particle energy band dispersion, with chemical potentials $\mu^{\ell}$.
While spin-orbit interactions\cite{Rashba1960}  are in general important for the hole bands in GaAs, 
here they can be neglected because narrow wells suppress Rashba and Dresselhaus interactions due to the large light-hole 
heavy-hole splitting.  Furthermore, the electric fields across the well here are small and the hole densities of interest here are low.  
The $V^{\ell \ell'}_{\mathbf{k}-\mathbf{k}'}$ are the bare Coulomb interaction 
potentials between electrons and holes confined in their finite width wells $\ell$ and $\ell'$.
The full expressions for $V^{\ell \ell'}_{\mathbf{k}-\mathbf{k}'}$ are found in Sec.\ S1 of 
the supplementary material.\cite{SupplementalGaAs1}

The mean-field equations at zero temperature 
for the superfluid gap $\Delta_{\mathbf{k}}$ and the average chemical potential 
$\mu=(\mu^{e}+\mu^{h})/2$, for equal electron and hole densities $n$ are,  
\begin{eqnarray}
	\Delta_{\mathbf{k}}
	&=& - \frac{1}{A}\sum_{\mathbf{k}'} V^{sc}_{\mathbf{k}-\mathbf{k}'} \frac{\Delta_{\mathbf{k}'} }{2 E_{\mathbf{k}'}} \ ,
	\label{Delta-eqn} 
	\\
	n &=& \frac{2}{A}\sum_{\mathbf{k}} (v_{\mathbf{k}})^2  \ ,
	\label{n-eh}
\end{eqnarray}
where $A$ is the sample surface area, $E_{\mathbf{k}} = \sqrt{\xi_{\mathbf{k}}^2 + \Delta_{\mathbf{k}}^{2}}$  
with  
$\xi_{\mathbf{k}}=\frac{1}{2}(\xi^{\mathrm{e}}_{\mathbf{k}} + \xi^{\mathrm{h}}_{\mathbf{k}})$, 
and the Bogoliubov coherence factors are,
\begin{equation}
	u_{\mathbf{k}}^2 = \frac{1}{2}\left(1+\frac{\xi_{\mathbf{k}}}{E_{\mathbf{k}}}\right);\ \ \ \ \ \ \ \ 
	v_{{\mathbf{k}}}^2 = \frac{1}{2}\left(1-\frac{\xi_{\mathbf{k}}}{E_{\mathbf{k}}}\right)\ .
	\label{uandv}
\end{equation}
$ V^{sc}_{\mathbf{q}}$ is the static screened electron-hole Coulomb interaction in the superfluid state
for momentum transfer $\mathbf{q}$, 
evaluated within the Random Phase Approximation (RPA) for electrons and holes of unequal masses. 
The superfluid energy gap $\Delta$ near the Fermi surface 
blocks excitations from the Fermi sea with energies less than $\Delta$.
This weakens the effect of screening, since low-lying excitations are those  
needed to screen the long-range Coulomb interactions.   The small-$\mathbf{q}$ suppression of 
screening leads to strong electron-hole pairing peaked at small-$\mathbf{q}$, and this can lead to large superfluid gaps.
The expressions for $ V^{sc}_{\mathbf{q}}$ are given in Sec.\ S2 of the supplementary material.\cite{SupplementalGaAs1} 

A comparison of the good agreement of zero-T superfluid properties for a double layer electron-hole system calculated using the present mean-field RPA approach,\cite{Neilson2014} with the corresponding results calculated using diffusion quantum Monte Carlo,\cite{LopezRios2018} indicates that the present RPA approach should be a quantitatively good approximation.

For a quasi-two-dimensional system, the superfluid transition is a topological transition associated with the Berezinskii-Kosterlitz-Thouless 
(BKT) transition temperature\cite{Kosterlitz1973} which depends only on the superfluid stiffness $\rho_s(T)$:\cite{Botelho2006}  
$T^{BKT} =  (\pi/2)\rho_s(T^{BKT})$.  
Provided $T^{BKT} \ll \Delta(T\!\!=\!\!0)$, which is the case here, 
$T^{BKT}  \simeq n \left[\pi {\hbar^2}/{(8(m^\star_e+m^\star_h))}\right]$. 
%

%
\begin{figure}[t]
	\includegraphics[angle=0,width=1\columnwidth] {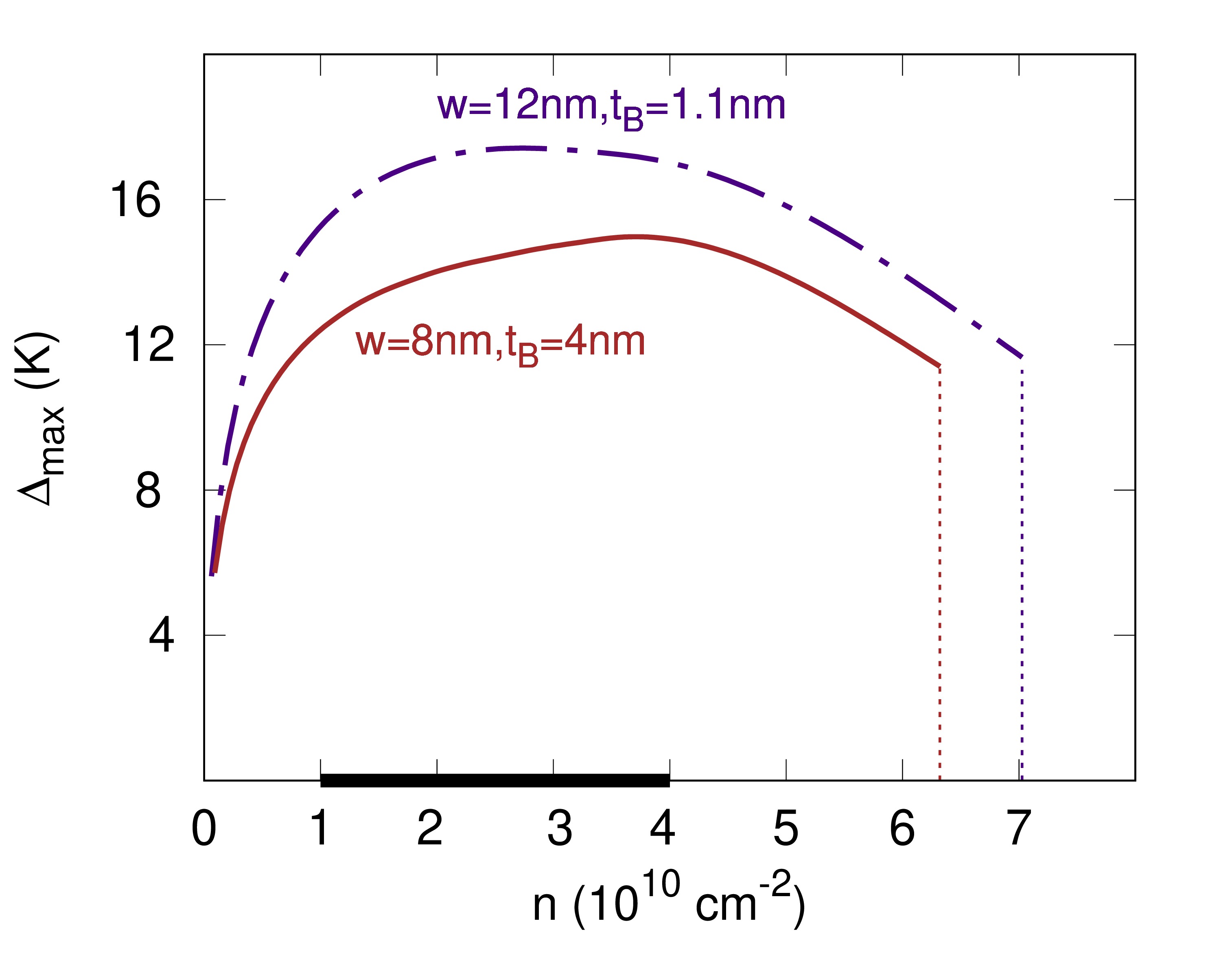}
	\caption{ (Color online).  $\Delta_{\mathrm{max}}$ is the maximum value of the zero-$T$ momentum-dependent superfluid gap 
		as a function of  $n$, the equal electron and hole densities. 
		Solid red line: $w=8$ nm and $t_B=4$ nm (sample: Refs.\ \citenum{Butov2002,High2012,Anankine2017}); 	
		dash-dot purple line: $w=12$ nm and $t_B=1.1$ nm (sample: Ref.\ \citenum{Timofeev2007}) .
		The horizontal black bar indicates the density range over which anomalous behavior was observed 
		in Refs.\ \citenum{Butov2002,High2012,Anankine2017}.  
	}
	\label{Fig.2}
\end{figure}
Figure \ref{Fig.2} shows $\Delta_{\mathrm{max}}$, the maximum of the zero-$T$ momentum-dependent gap $\Delta_{\bf k}$  
determined from Eq.\ \eqref{Delta-eqn}, as a function of  $n$, the equal electron and hole densities
for the GaAs heterostructures used for the optical observations\cite{Butov2002,High2012,Anankine2017,Timofeev2007}. 
Figure \ref{Fig.2} illustrates  that electron-hole superfluidity is a low-density phenomenon:  
at higher densities, strong screening greatly weakens the electron-hole coupling, leading  
to $\Delta_{\mathrm{max}}$ of, at most, only a few mK\cite{Lozovik2012, Perali2013}.  
In a real system, such small $\Delta_{\mathrm{max}}$ would be destroyed by disorder.  
As the density is reduced to $n_0$, the onset density, $\Delta_{\mathrm{max}}$ jumps to energies $\gtrsim 5$ K.  
This is a self-consistent effect since large gaps weaken the screening. 
In Fig.\ \ref{Fig.2}, the onset densities are relatively high, $n_0\sim 6$ - $7\times 10^{10}$ cm$^{-2}$, thanks to the narrow wells and barriers 
of the samples in these experiments. For narrower wells and barriers, the average separation between electrons and holes is smaller which strengthens the pairing interactions. The range of densities at which anomalous behavior was observed 
in Refs.\ \citenum{Butov2002,High2012,Anankine2017}
is indicated as the green bar on the figure.  We note that this lies within the density range for which we predict superfluidity.     
This adds independent credence that the observed anomalous behavior is indeed associated with superfluidity.
We find the maximum Berezinskii-Kosterlitz-Thouless transition temperature is a few Kelvin. 

In Sec.\  S3 of the supplementary material\cite{SupplementalGaAs1} we discuss the property that the superfluidity in the GaAs system is nearly always confined within the crossover regime that separates the BCS regime from the BEC regime.  This is due to the strong screening that kills superfluidity in the weakly-interacting BCS regime, and to the large electron-hole mass difference in GaAs that impedes the system at low densities from entering the strongly-interacting BEC regime.ring the strongly-interacting BEC regime at low densities.

%
\begin{figure}[t]
	\includegraphics[angle=0,width=1\columnwidth] {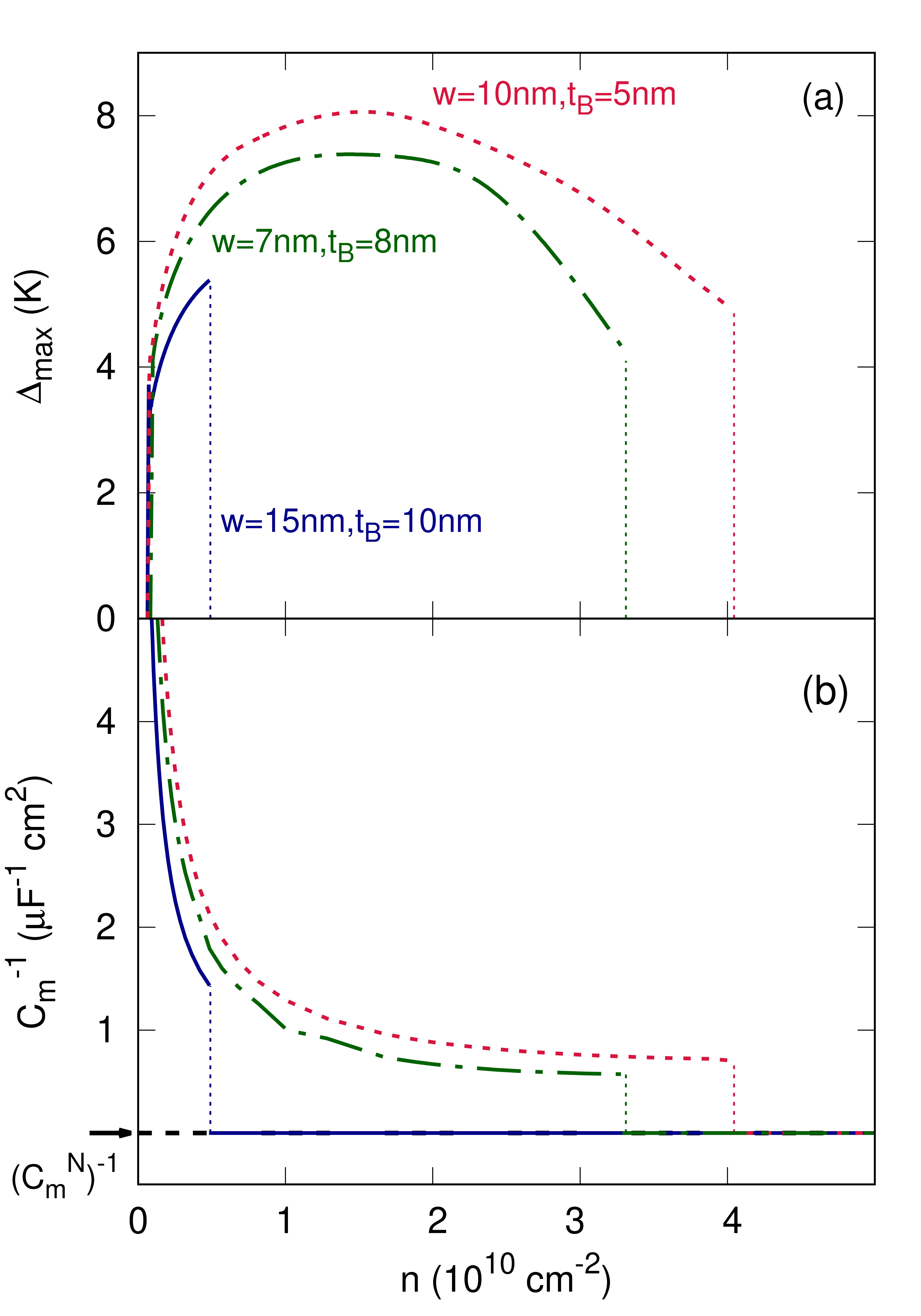}
	\caption{
		(a) $\Delta_{\mathrm{max}}$ as a function of $n$. 
		Solid blue line: $w=15$ nm and $t_B=10$ nm (sample: Ref.\ \citenum{Croxall2008});  
		dash-dot green line:  $w=7$ nm and $t_B=8$ nm; 
		dotted red line: $w=10$ nm and $t_B=5$ nm.
		(b) Corresponding inverse of total capacitance $C_m^{-1}$ as a function of density $n$. 
		The inverse of total capacitance in the normal state, $\left(C_m^N\right)^{-1}$, is indicated.     
	}
	\label{Fig.3}
\end{figure}

As we have discussed, to get high enough mobilities to avoid localization and allow transport studies, 
the wells and the barriers need to be wider for the transport drag measurements\cite{Croxall2008,Seamons2009} 
compared with the samples for the optical measurements.
We see in Fig.\ \ref{Fig.3}(a), showing $\Delta_{\mathrm{max}}$ as a function of  $n$, that for 
well widths $w\!\!=\!\!15$ nm and barrier thickness $t_B\!\!=\!\!10$ nm, the 
onset density $n_0\sim 0.5\times 10^{10}$ cm$^{-2}$, is an order of magnitude smaller than for the optical measurements.  
Since the lowest density attained in the drag experiments, was $n\gtrsim 4\times 10^{10} \gg n_0$, we conclude that  
the anomalous behavior reported in the drag experiments is most probably not an indication of a superfluid transition.  

Figure \ref{Fig.3}(a) shows the onset density could be markedly increased 
to reach the minimum densities attained in Ref.\ \citenum{Croxall2008}, 
by relatively minor reductions in $w$ and $t_B$.
However, interface roughness scattering increases rapidly as the well is made narrower,
resulting in samples with very low mobilities.  
Nevertheless, even if no macroscopically connected superfluid remained, 
superfluidity may well survive in pockets randomly distributed along the quantum wells.   
Such pockets of superfluidity could be detected using capacitance spectroscopy.\cite{Du2017}

In capacitance spectroscopy, 
a low frequency ac voltage is delivered to a top gate with the quantum wells grounded.  The 
total capacitance  $C_m = (C_g^{-1} + C_Q^{-1})^{-1}$ between the gate and quantum wells is measured.  
$C_g$ is the classical geometry capacitance per unit area which depends only on the sample structure.  
$C_Q= e^2 \partial n/\partial \mu$ is the quantum capacitance and is proportional to the density of states.
For a two-dimensional system in the normal state $C^N_Q= (1/A) [e^2 m^*/(\pi \hbar^2)]$,      
inversely proportional to the sample area. 
For a pocket of superfluidity of area $A^\prime\leq A$, the quantum capacitance is, 
\begin{equation}
	C^S_Q= \frac{1}{A^\prime} \left[e^2 \sum_{\mathbf{k}} \delta(\mu-\epsilon_F) + 4 e^2 \sum_{\mathbf{k}} \frac{u_{\mathbf{k}}^2 v_{\mathbf{k}}^2}{E_k}\right]\ .
\end{equation} 
$C^S_Q < C_Q^N$, because of the gap in the low-lying energy spectrum.\cite{Yang1997,Du2017}  

Figure\ \ref{Fig.3}(b) shows 
the inverse of the total capacitance $C_m^{-1}$ as a function of density for homogeneous systems.  
From Fig.\ \ref{Fig.3}(a), the onset density for superfluidity for well widths $w= 10$ nm and barrier thickness $t_B=5$ nm,  is $n_0\sim 4\times 10^{10}$ cm$^{-2}$.    
$C_m^N$ is the corresponding total capacitance in the normal state.  
The onset of  superfluidity is characterised by a jump in  $C_m^{-1}$ at $n_0$, and $C_m^{-1}$ monotonically increases as  
the density is further decreased.
For the inhomogeneous system with pockets of superfluid of total area $A^\prime$, the behavior would be similar, 
but the jump in  $C_m^{-1}$ at $n_0$ would be reduced by an amount proportional to $(A'/A)$.

\begin{table}[t]
	\caption{A: Double quantum well structure from Ref.\ \citenum{Croxall2008}; 
		B: same structure, calculated without band bending;
		C: same structure, calculated for the same $d_c$ but with the width of the quantum wells neglected; 
		D: A different double quantum well structure with the same $d_c$ as A.\\
		Columns: $w$ quantum well widths, 
		$t_B$ barrier thickness, 
		$d_c$ distance between centers of the two wells,
		$d_p$ distance between the peaks of the electron and hole density distributions,  
		$n_0$ superfluid onset density with $r_0$ the corresponding average inter-electron spacing, 
		$\bar{\Delta}$ the maximum value of $\Delta_{max}$ across all densities, and 
		$T^{BKT}$  maximum superfluid transition temperature.}
	\label{table}
	\begin{tabular}{p {0.02\textwidth} p {0.045\textwidth} p {0.045\textwidth}  p {0.045\textwidth} p {0.045\textwidth} 
			p {0.058\textwidth} p {0.05\textwidth} p {0.045\textwidth} p {0.045\textwidth}}            
		\hline	\hline
		& \textbf{w} (nm) & \textbf{t}$\bm{_B}$ (nm) & \textbf{d}$\bm{_c}$  (nm)  & \textbf{d}$\bm{_p}$ (nm) 
		& \textbf{n}$\bm{_0}$ \!\!\!\!\!\!($10^{10}$ cm$^{-2}$)  
		& 	$\bm{r_0/}$\textbf{d}$\bm{_p}$ & $\bm{\bar{\Delta}}$ (K)  & 	$\bm{T^{BKT}}$  (K)    \\ 
		\hline
		\textbf{A} &		15 & 10 & 25 & 27 & 0.5  & 3.0 & 5.2 & 0.1 \\  
		\textbf{B} &		15 & 10 & 25 & 25 & 0.8  & 2.5 & 6.2 & 0.3  \\  
		\textbf{C} &		15 & 10 & 25 & 25 & 1.0  & 2.3 & 8.3 & 0.4   \\  
		\textbf{D} &		10 & 15 & 25 & 29 &  0.4 & 3.1 & 4.0 & 0.1  \\   
		\hline\hline	  
	\end{tabular}
\end{table}

Table \ref{table} shows the effects on the superfluid properties of band bending and the finite width of the quantum wells
for samples from Ref.\ \citenum{Croxall2008}.  We saw in Fig.\ \ref{Fig.1} that band bending 
pushes the peaks of the electron and hole density distributions ($d_p$) further apart 
than the distance between the centers of the wells ($d_c$).  The effect of this in weakening the superfluidity
can be seen by comparing rows A and B.  In row B, band bending has been neglected.    
The finite thickness of the wells also weakens the superfluidity. This can be seen by comparing row A with row C, 
calculated for the same $d_c$ but neglecting the well widths. For a fixed distance between the centers of the wells $d_c$, 
narrower wells with a thicker barrier also weaken the superfluidity, a combined effect of banding bending 
and the gate potentials.\cite{Zheng2016} This is seen by comparing rows A and D.  
The ratio $r_0/d_p$, where $r_0$ is the average spacing of the electrons at the superfluid onset density $n_0$, is a useful indicator
of the effect of the heterostructure parameters on $n_0$.  The table shows that $n_0$ occurs for a value of the ratio
$r_0/d_p\sim 2.5$ - $3$.  For smaller $r_0$, the screening is too strong and the superfluidity cannot overcome the screening.  


The unusually large effective mass difference between electrons and holes in GaAs makes experimental realization of superfluidity in GaAs double 
quantum wells a particularly worthwhile goal, likely to reveal intriguing new physics.  While in principle this physics can also be investigated with 
ultracold atoms of different masses, the exotic superfluid phases are predicted for the ultracold atom system only below nanoKelvin temperatures, 
whereas for GaAs the transition temperatures are a few Kelvin.  The primary reason it has proved so difficult to observe superfluidity in GaAs is that the 
phenomenon only occurs at low densities, due to strong screening at higher densities of the long-range Coulomb pairing interactions.  Screening kills 
superfluidity above an onset density.  To generate superfluidity in GaAs at accessible experimental densities requires narrow quantum wells and thin 
barriers that have been impractical for transport experiments because of the very low mobility of the samples.  However, we show that inhomogeneous 
pockets of superfluidity could be detected in samples of low mobility using standard capacitance measurement techniques.  Finally, our calculations 
confirm that superfluid condensation of optically generated electron-hole pairs is indeed feasible with existing experimental samples.  

{\bf Acknowledgments.}
We thank Kanti Das Gupta, Fran\c{c}ois Dubin, Ugo Siciliani de Cumis, Michele Pini, and Joanna Waldie for illuminating discussions.
This work was partially supported by the Flemish Science Foundation (FWO-Vl),
and the Australian Government through the Australian Research Council Centre of Excellence 
in Future Low-Energy Electronics (Project No. CE170100039).
\newpage
\section{SUPPLEMENTARY MATERIAL \\
	Experimental conditions for observation of electron-hole superfluidity in GaAs heterostructures}

\maketitle
\newcommand{\beginsupplement}{%
	\setcounter{table}{0}
	\renewcommand{\thetable}{S\arabic{table}}%
	\setcounter{figure}{0}
	\renewcommand{\thefigure}{S\arabic{figure}}%
	\setcounter{section}{0}
	\renewcommand{\thesection}{S\arabic{section}}%
	\setcounter{equation}{0}
	\renewcommand{\theequation}{S\arabic{equation}}%
}
\beginsupplement
\section{Bare Coulomb interactions}

The $V^{\ell \ell'}_{\mathbf{k}-\mathbf{k}'}$ are the bare Coulomb interaction 
potentials between electrons and holes confined in their wells $\ell$ and $\ell'$,
\begin{eqnarray}
	V^{\ell \ell}_{\mathbf{k}-\mathbf{k}'}
	&=& \frac{2\pi e^2}{\epsilon}\frac{1}{|\mathbf{k}-\mathbf{k}'|}
	F^{\ell\ell}_{\mathbf{k}-\mathbf{k}'} \ ; \nonumber \\
	V^{\ell \neq \ell'}_{\mathbf{k}-\mathbf{k}'}
	&=& -\frac{2\pi e^2}{\epsilon}\frac{e^{-d|\mathbf{k}-\mathbf{k}'|}}{|\mathbf{k}-\mathbf{k}'|} 
	F^{\ell\ell'}_{\mathbf{k}-\mathbf{k}'} \ .
	\label{bare_interactions}
\end{eqnarray}
$V^{\ell \ell}_{\mathbf{k}-\mathbf{k}'}$ is the bare intralayer interaction 
and $V^{\ell \neq \ell'}_{\mathbf{k}-\mathbf{k}'}$ the bare interlayer interaction.  
We take the dielectric constant for GaAs and Al$_{0.9}$Ga$_{0.1}$As as $\epsilon = 12.9$. We consider only wells of equal width $w$, separated by a Al$_{0.9}$Ga$_{0.1}$As insulating barrier of thickness $t_B$. 
We express lengths in units of the effective Bohr radius,  
$a_{B}^{*} = \hbar^2 4\pi \epsilon_0\epsilon/(m^* e^2) = 12.5$ nm for GaAs, and  
energies in effective Rydbergs, Ry$^{*}=e^2/(2a_{B}^{*}) = 4.5$ meV $=52$ K for GaAs.   
$m^*$ is the reduced effective mass.  

The form-factors in Eq.\ \eqref{bare_interactions}, 
\begin{equation}
	\!F^{\ell\ell'}_\mathbf{q} \!\!\!=\!\!\! \int^\infty_{-\infty}\!\!\! \textrm{d}z 
	\int^\infty_{-\infty}\!\!\! \textrm{d}z' |\phi_\ell(z)|^2 
	|\phi_{\ell'}(z')|^2  \exp(-q|z-z'|) ,
	\label{formfactor}
\end{equation} 
account for the confinement of the electrons and holes in their finite-width quantum wells.  
$\phi_\ell(z)$ is the wave-function of the electron or hole (see Fig.\ 1 of main manuscript).
The $\phi_\ell(z)$ are evaluated numerically using a self-consistent Poisson-Schr\"{o}dinger solver.\cite{Tan1990}  
They are sensitive to the voltages on the front and back gates that are   
used to independently tune the populations in the well.
The $\phi_\ell(z)$ are also sensitive to any external interlayer bias.
In the optical measurements, an electric field across the barrier spatially separates  the  
optically generated electrons and holes into opposite wells, while in the transport drag measurements   
a bias is used to offset the different electrochemical potentials 
of the holes and electrons.\cite{Croxall2008,Seamons2009,Sivan1992}\\

\section{Screened Coulomb interaction in  superfluid state}

In the normal state, the static screened electron-hole Coulomb interaction evaluated within the Random Phase Approximation (RPA)  for electrons and holes of unequal masses is,\cite{Flensberg1995} 
%
\begin{equation}
	V^{sc}_{\mathbf{q}} \!=\! \frac{V^{eh}_{\mathbf{q}}}
	{ [1\!-\!V^{ee}_{\mathbf{q}}\Pi_0^e(\mathbf{q})] [1\!-\!V^{hh}_{\mathbf{q}}\Pi_0^h(\mathbf{q})]
		-[V^{eh}_{\mathbf{q}}]^2\Pi_0^e(\mathbf{q})\Pi_0^h(\mathbf{q})} 
	\label{V^{eh}_norm}
\end{equation}
%
where 
\begin{equation}
	\Pi_0^\ell(\mathbf{q}) = 2\sum_{\mathbf{k}}
	\frac{f_{FD}(\xi_{\mathbf{k}-\mathbf{q}}^{\ell})-f_{FD}(\xi_{\mathbf{k}}^{\ell})}
	{\xi_{\mathbf{k}-\mathbf{q}}^{\ell}-\xi_{\mathbf{k}}^{\ell}}\ 
	\label{Pi_0^n}
\end{equation}
is the Lindhard particle-hole polarization in well $\ell$.  
$f_{FD}(x)$ 
is the  zero-$T$ Fermi-Dirac distribution function. 

\begin{widetext}
	In the superfluid state, the expression for the static RPA electron-hole Coulomb interaction $V^{sc}_{\mathbf{q}}$ 
	is different,\cite{Iskin2007,Lozovik2012,Sodemann2012} 
	\begin{equation}
		V^{sc}_{\mathbf{q}}  =\frac{V^{eh}_\mathbf{q} +\Pi_{a}(\mathbf{q})[V^{ee}_\mathbf{q} V^{hh}_\mathbf{q}-(V^{eh}_\mathbf{q} )^2]} 
		{ 1-\Pi_n(\mathbf{q})[V^{ee}_\mathbf{q}+ V^{hh}_\mathbf{q}]+ 2V^{eh}_\mathbf{q}\Pi_a(\mathbf{q})
			+[V^{ee}_\mathbf{q} V^{hh}_\mathbf{q}-(V^{eh}_\mathbf{q} )^2][\Pi_n^2(\mathbf{q})-\Pi_a^2(\mathbf{q})]} \ ,
		\label{eq:VeffSF}
	\end{equation}
	where $\Pi_n(\mathbf{q})$ is the normal polarizability, modified from the polarizabilities in 
	Eq.\ \eqref{Pi_0^n} for the normal state by the presence of the superfluid gap in the energy spectrum, 
	\begin{equation}
		\begin{aligned}
			\Pi_{n}(\boldsymbol{q})=2 \sum_{\boldsymbol{k}} \bigg[
			&u^2_{\boldsymbol{k}}u^2_{\boldsymbol{k}-\boldsymbol{q}}
			\frac{f_{FD}(E^{+}_{\boldsymbol{k}-\boldsymbol{q}})
				-f_{FD}(E^+_{\boldsymbol{k}})}{E^{+}_{\boldsymbol{k}-\boldsymbol{q}}-E^+_{\boldsymbol{k}}}
			+v^2_{\boldsymbol{k}}v^2_{\boldsymbol{k}-\boldsymbol{q}}
			\frac{f_{FD}(E^{-}_{\boldsymbol{k}-\boldsymbol{q}})-
				f_{FD}(E^-_{\boldsymbol{k}})}{E^{-}_{\boldsymbol{k}-\boldsymbol{q}}-E^-_{\boldsymbol{k}}}\\
			&-u^2_{\boldsymbol{k}}v^2_{\boldsymbol{k}-\boldsymbol{q}}
			\frac{1-f_{FD}(E^{-}_{\boldsymbol{k}-\boldsymbol{q}})
				-f_{FD}(E^+_{\boldsymbol{k}})}{E^{-}_{\boldsymbol{k}-\boldsymbol{q}}+E^+_{\boldsymbol{k}}}
			-v^2_{\boldsymbol{k}}u^2_{\boldsymbol{k}-\boldsymbol{q}}
			\frac{1-f_{FD}(E^{+}_{\boldsymbol{k}-\boldsymbol{q}})
				-f_{FD}(E^-_{\boldsymbol{k}})}{E^{+}_{\boldsymbol{k}-\boldsymbol{q}}+E^{-}_{\boldsymbol{k}}} \bigg] \ ,
			\label{pi^n_super}
		\end{aligned}
	\end{equation}
	with 
	$E^{\pm}_{\mathbf{k}} = E_{\mathbf{k}} \pm \delta \xi_{\mathbf{k}}$  
	and 
	$\delta \xi_{\mathbf{k}} =\frac{1}{2}(\xi^{\mathrm{e}}_{\mathbf{k}} - \xi^{\mathrm{h}}_{\mathbf{k}})$, and 
	$\Pi_a(\mathbf{q})$ is the anomalous polarizability,\cite{Lozovik2012,Perali2013} 
	\begin{equation}
		\begin{aligned}		
			\Pi_{a}(\boldsymbol{q})=2 \sum_{\boldsymbol{k}}  u_{\boldsymbol{k}} v_{\boldsymbol{k}
				-\boldsymbol{q}}  v_{\boldsymbol{k}} u_{\boldsymbol{k}-\boldsymbol{q}} 
			\bigg[&\frac{f_{FD}(E^{+}_{\boldsymbol{k}-\boldsymbol{q}})-f_{FD}(E^+_{\boldsymbol{k}})}
			{E^{+}_{\boldsymbol{k}-\boldsymbol{q}}-E^+_{\boldsymbol{k}}}
			-\frac{f_{FD}(E^{-}_{\boldsymbol{k}-\boldsymbol{q}})-f_{FD}(E^-_{\boldsymbol{k}})}
			{-E^{-}_{\boldsymbol{k}-\boldsymbol{q}}+E^-_{\boldsymbol{k}}}\\
			&+\frac{1-f_{FD}(E^{-}_{\boldsymbol{k}-\boldsymbol{q}})-f_{FD}(E^+_{\boldsymbol{k}})}
			{E^{-}_{\boldsymbol{k}-\boldsymbol{q}}+E^+_{\boldsymbol{k}}}
			+\frac{1-f_{FD}(E^{+}_{\boldsymbol{k}-\boldsymbol{q}})
				-f_{FD}(E^-_{\boldsymbol{k}})}{E^{+}_{\boldsymbol{k}-\boldsymbol{q}}+E^{-}_{\boldsymbol{k}}}
			\bigg] \ .
			\label{pi^a_super}
		\end{aligned}
	\end{equation}
	
\end{widetext}
Equations \ \ref{eq:VeffSF}-\ref{pi^a_super} include the effect of the unequal masses. The full polarization is $\Pi_0(\mathbf{q}) = \Pi_{n}(\boldsymbol{q})+\Pi_{a}(\boldsymbol{q})$. In the normal state $\Pi_{a}(\boldsymbol{q})=0$. In the superfluid state $\Pi_{a}({0})= -\Pi_{n}({0})$ so $\Pi_0(\mathbf{0}) =0$, and $\Pi_0(\mathbf{q})$ is suppressed for small $\boldsymbol{q}$ by the presence of the gap in the energy spectrum.\cite{Gortel1996,Bistritzer2008} 

\section{Condensate fraction, chemical potential and intrapair correlation length}

\begin{figure}[H]
	\includegraphics[angle=0, width=1\columnwidth] {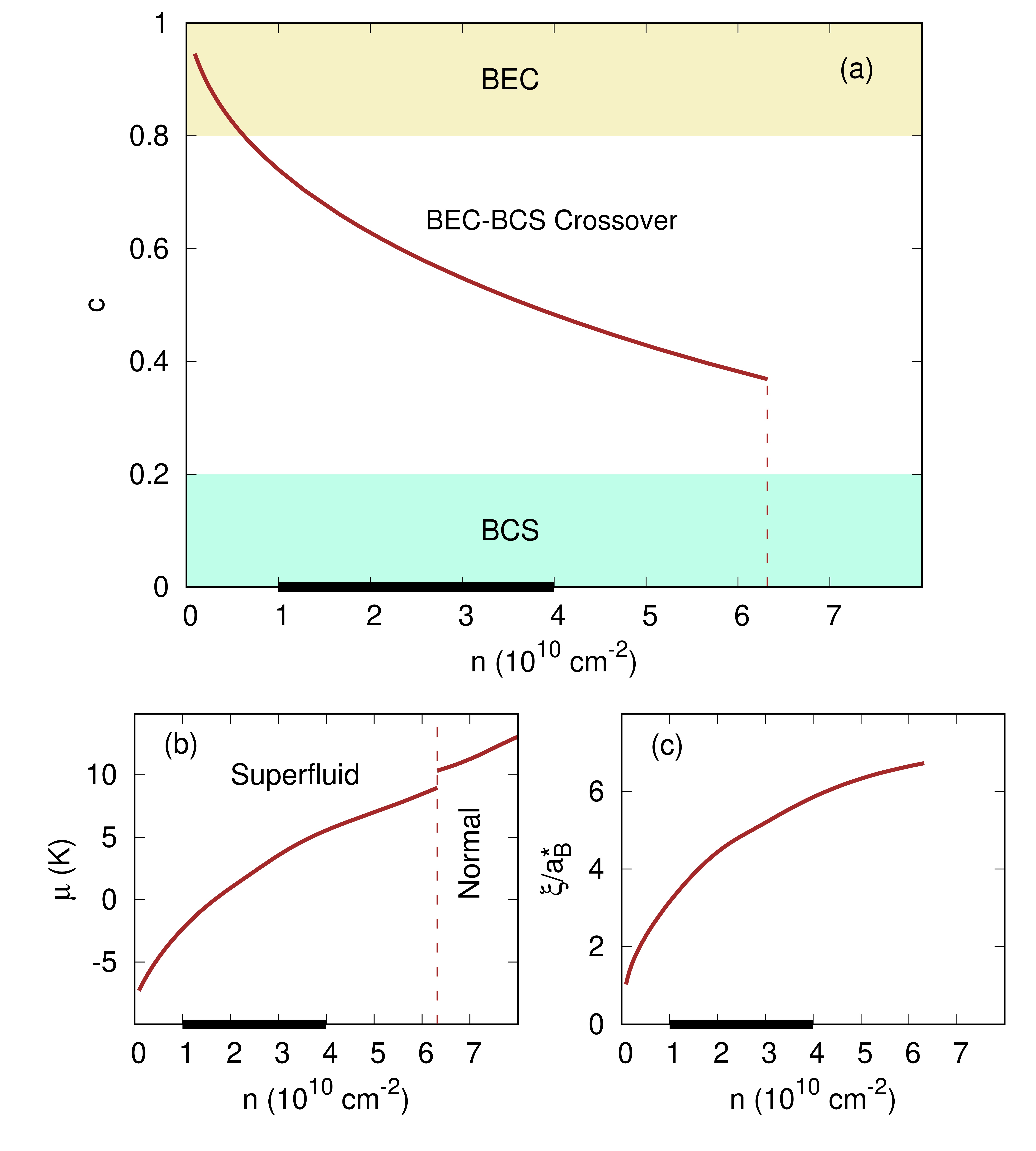}
	\caption{  (a) Condensate fraction $c$ 
		as a function of equal electron and hole densities $n$,  
		for a double quantum well in GaAs with well widths $w=8$ nm and barrier thickness $t_B=4$ nm (sample: Refs.\ \citenum{Butov2002,High2012,Anankine2017}). 	
		The superfluid onset density is $n_0=6.3\times 10^{10}$ cm$^{-2}$, indicated by vertical dashed line.
		(b) Corresponding chemical potential $\mu$.  
		(c) Corresponding intra pair  correlation length $\xi$. 
		$a_B^\star$ is the effective Bohr radius. The horizontal black bar indicates the density range over which anomalous behavior was observed 
		in Refs.\ \citenum{Butov2002,High2012,Anankine2017}.
	}
	\label{Fig.1S}
\end{figure}

The superfluid BCS, BCS-BEC crossover and BEC regimes can be characterized by the condensate fraction $c$, 
the fraction of carriers bound in pairs.\cite{Salasnich2005,Guidini2014}  
For $c \leq 0.2$ the superfluid condensate is in the BCS regime, for $0.2 <c < 0.8$  in the crossover regime, and 
for $c \geq 0.8$ in the BEC regime.   In Fig.\ \ref{Fig.1S}(a) we see the condensate fraction $c$ as a function of density, for a sample with well widths $w=8$ nm and barrier thickness $t_B=4$ nm.     
The figure shows that the superfluidity is essentially always confined to the crossover regime.
At the superfluid onset density $n_0\sim 6.3 \times 10^{10}$ cm$^{-2}$, $c$  drops sharply from $c= 0.36$ to exponentially small values  (see Ref.\ \citenum{Neilson2014} and the discussion of the behavior of $\Delta_{max}$ near $n_0$ in the main manuscript).  A value of $0.36$  implies that when the superfluidity starts it is in the crossover region. Strong screening kills superfluidity in the BCS regime. 

It is also difficult for the system to enter the BEC regime at low densities, 
a consequence of the large mass difference:  even for pairs compact on the scale of the interparticle spacing, the electrons and holes in the pairs  
can still respond and screen an electric field.  

Fig.\ \ref{Fig.1S}(b) and (c) show as a function of density, the chemical potential $\mu$ and  
the intra pair correlation length $\xi$, defined as the radius of an isolated electron-hole bound pair.\cite{Pistolesi1996} 
Above the onset density, the system is in the normal state and $\mu$ is equal to the Fermi energy.
$\mu$ drops at the onset density, but it only reaches   
negative values, corresponding to the deep BEC regime, at extremely low density.  
Similarly, $\xi$ only drops to  the effective Bohr radius $a_B^\star$ at low density.  

\makeatletter

\begin{thebibliography}{52}%
	\makeatletter
	\providecommand \@ifxundefined [1]{%
		\@ifx{#1\undefined}
	}%
	\providecommand \@ifnum [1]{%
		\ifnum #1\expandafter \@firstoftwo
		\else \expandafter \@secondoftwo
		\fi
	}%
	\providecommand \@ifx [1]{%
		\ifx #1\expandafter \@firstoftwo
		\else \expandafter \@secondoftwo
		\fi
	}%
	\providecommand \natexlab [1]{#1}%
	\providecommand \enquote  [1]{``#1''}%
	\providecommand \bibnamefont  [1]{#1}%
	\providecommand \bibfnamefont [1]{#1}%
	\providecommand \citenamefont [1]{#1}%
	\providecommand \href@noop [0]{\@secondoftwo}%
	\providecommand \href [0]{\begingroup \@sanitize@url \@href}%
	\providecommand \@href[1]{\@@startlink{#1}\@@href}%
	\providecommand \@@href[1]{\endgroup#1\@@endlink}%
	\providecommand \@sanitize@url [0]{\catcode `\\12\catcode `\$12\catcode
		`\&12\catcode `\#12\catcode `\^12\catcode `\_12\catcode `\%12\relax}%
	\providecommand \@@startlink[1]{}%
	\providecommand \@@endlink[0]{}%
	\providecommand \url  [0]{\begingroup\@sanitize@url \@url }%
	\providecommand \@url [1]{\endgroup\@href {#1}{\urlprefix }}%
	\providecommand \urlprefix  [0]{URL }%
	\providecommand \Eprint [0]{\href }%
	\providecommand \doibase [0]{http://dx.doi.org/}%
	\providecommand \selectlanguage [0]{\@gobble}%
	\providecommand \bibinfo  [0]{\@secondoftwo}%
	\providecommand \bibfield  [0]{\@secondoftwo}%
	\providecommand \translation [1]{[#1]}%
	\providecommand \BibitemOpen [0]{}%
	\providecommand \bibitemStop [0]{}%
	\providecommand \bibitemNoStop [0]{.\EOS\space}%
	\providecommand \EOS [0]{\spacefactor3000\relax}%
	\providecommand \BibitemShut  [1]{\csname bibitem#1\endcsname}%
	\let\auto@bib@innerbib\@empty
	\bibitem [{\citenamefont {Bartenstein}\ \emph {et~al.}(2004)\citenamefont
		{Bartenstein}, \citenamefont {Altmeyer}, \citenamefont {Riedl}, \citenamefont
		{Jochim}, \citenamefont {Chin}, \citenamefont {Denschlag},\ and\
		\citenamefont {Grimm}}]{Bartenstein2004}%
	\BibitemOpen
	\bibfield  {author} {\bibinfo {author} {\bibfnamefont {M.}~\bibnamefont
			{Bartenstein}}, \bibinfo {author} {\bibfnamefont {A.}~\bibnamefont
			{Altmeyer}}, \bibinfo {author} {\bibfnamefont {S.}~\bibnamefont {Riedl}},
		\bibinfo {author} {\bibfnamefont {S.}~\bibnamefont {Jochim}}, \bibinfo
		{author} {\bibfnamefont {C.}~\bibnamefont {Chin}}, \bibinfo {author}
		{\bibfnamefont {J.~Hecker}\ \bibnamefont {Denschlag}}, \ and\ \bibinfo
		{author} {\bibfnamefont {R.}~\bibnamefont {Grimm}},\ }\bibfield  {title}
	{\enquote {\bibinfo {title} {Crossover from a molecular {B}ose-{E}instein
				{C}ondensate to a degenerate {F}ermi gas},}\ }\href {\doibase
		10.1103/PhysRevLett.92.120401} {\bibfield  {journal} {\bibinfo  {journal}
			{Phys. Rev. Lett.}\ }\textbf {\bibinfo {volume} {92}},\ \bibinfo {pages}
		{120401} (\bibinfo {year} {2004})}\BibitemShut {NoStop}%
	\bibitem [{\citenamefont {Regal}\ \emph {et~al.}(2004)\citenamefont {Regal},
		\citenamefont {Greiner},\ and\ \citenamefont {Jin}}]{Regal2004}%
	\BibitemOpen
	\bibfield  {author} {\bibinfo {author} {\bibfnamefont {C.~A.}\ \bibnamefont
			{Regal}}, \bibinfo {author} {\bibfnamefont {M.}~\bibnamefont {Greiner}}, \
		and\ \bibinfo {author} {\bibfnamefont {D.~S.}\ \bibnamefont {Jin}},\
	}\bibfield  {title} {\enquote {\bibinfo {title} {Observation of resonance
			condensation of fermionic atom pairs},}\ }\href {\doibase
	10.1103/PhysRevLett.92.040403} {\bibfield  {journal} {\bibinfo  {journal}
		{Phys. Rev. Lett.}\ }\textbf {\bibinfo {volume} {92}},\ \bibinfo {pages}
	{040403} (\bibinfo {year} {2004})}\BibitemShut {NoStop}%
\bibitem [{\citenamefont {Zwierlein}\ \emph {et~al.}(2004)\citenamefont
	{Zwierlein}, \citenamefont {Stan}, \citenamefont {Schunck}, \citenamefont
	{Raupach}, \citenamefont {Kerman},\ and\ \citenamefont
	{Ketterle}}]{Zwierlein2004}%
\BibitemOpen
\bibfield  {author} {\bibinfo {author} {\bibfnamefont {M.~W.}\ \bibnamefont
		{Zwierlein}}, \bibinfo {author} {\bibfnamefont {C.~A.}\ \bibnamefont {Stan}},
	\bibinfo {author} {\bibfnamefont {C.~H.}\ \bibnamefont {Schunck}}, \bibinfo
	{author} {\bibfnamefont {S.~M.~F.}\ \bibnamefont {Raupach}}, \bibinfo
	{author} {\bibfnamefont {A.~J.}\ \bibnamefont {Kerman}}, \ and\ \bibinfo
	{author} {\bibfnamefont {W.}~\bibnamefont {Ketterle}},\ }\bibfield  {title}
{\enquote {\bibinfo {title} {Condensation of pairs of fermionic atoms near a
			{F}eshbach resonance},}\ }\href {\doibase 10.1103/PhysRevLett.92.120403}
{\bibfield  {journal} {\bibinfo  {journal} {Phys. Rev. Lett.}\ }\textbf
	{\bibinfo {volume} {92}},\ \bibinfo {pages} {120403} (\bibinfo {year}
	{2004})}\BibitemShut {NoStop}%
\bibitem [{\citenamefont {Perali}\ \emph {et~al.}(2013)\citenamefont {Perali},
	\citenamefont {Neilson},\ and\ \citenamefont {Hamilton}}]{Perali2013}%
\BibitemOpen
\bibfield  {author} {\bibinfo {author} {\bibfnamefont {A.}~\bibnamefont
		{Perali}}, \bibinfo {author} {\bibfnamefont {D.}~\bibnamefont {Neilson}}, \
	and\ \bibinfo {author} {\bibfnamefont {A.~R.}\ \bibnamefont {Hamilton}},\
}\bibfield  {title} {\enquote {\bibinfo {title} {High-temperature
		superfluidity in double-bilayer graphene},}\ }\href {\doibase
10.1103/PhysRevLett.110.146803} {\bibfield  {journal} {\bibinfo  {journal}
	{Phys. Rev. Lett.}\ }\textbf {\bibinfo {volume} {110}},\ \bibinfo {pages}
{146803} (\bibinfo {year} {2013})}\BibitemShut {NoStop}%
\bibitem [{\citenamefont {Conti}\ \emph
	{et~al.}(2019{\natexlab{a}})\citenamefont {Conti}, \citenamefont {der Donck},
	\citenamefont {Perali}, \citenamefont {Peeters},\ and\ \citenamefont
	{Neilson}}]{Conti2019b}%
\BibitemOpen
\bibfield  {author} {\bibinfo {author} {\bibfnamefont {S.}~\bibnamefont
		{Conti}}, \bibinfo {author} {\bibfnamefont {M.~Van}\ \bibnamefont {der
			Donck}}, \bibinfo {author} {\bibfnamefont {A.}~\bibnamefont {Perali}},
	\bibinfo {author} {\bibfnamefont {F.~M.}\ \bibnamefont {Peeters}}, \ and\
	\bibinfo {author} {\bibfnamefont {D.}~\bibnamefont {Neilson}},\ }\href@noop
{} {\enquote {\bibinfo {title} {A doping-dependent switch from one- to
			two-component superfluidity at temperature above 100{K} in coupled
			electron-hole {V}an der {W}aals heterostructures},}\ } (\bibinfo {year}
{2019}{\natexlab{a}}),\ \Eprint {http://arxiv.org/abs/1909.03411}
{arXiv:1909.03411 [cond-mat.supr-con]} \BibitemShut {NoStop}%
\bibitem [{\citenamefont {Burg}\ \emph {et~al.}(2018)\citenamefont {Burg},
	\citenamefont {Prasad}, \citenamefont {Kim}, \citenamefont {Taniguchi},
	\citenamefont {Watanabe}, \citenamefont {MacDonald}, \citenamefont
	{Register},\ and\ \citenamefont {Tutuc}}]{Burg2018}%
\BibitemOpen
\bibfield  {author} {\bibinfo {author} {\bibfnamefont {G.~W.}\ \bibnamefont
		{Burg}}, \bibinfo {author} {\bibfnamefont {N.}~\bibnamefont {Prasad}},
	\bibinfo {author} {\bibfnamefont {K.}~\bibnamefont {Kim}}, \bibinfo {author}
	{\bibfnamefont {T.}~\bibnamefont {Taniguchi}}, \bibinfo {author}
	{\bibfnamefont {K.}~\bibnamefont {Watanabe}}, \bibinfo {author}
	{\bibfnamefont {A.~H.}\ \bibnamefont {MacDonald}}, \bibinfo {author}
	{\bibfnamefont {L.~F.}\ \bibnamefont {Register}}, \ and\ \bibinfo {author}
	{\bibfnamefont {E.}~\bibnamefont {Tutuc}},\ }\bibfield  {title} {\enquote
	{\bibinfo {title} {Strongly enhanced tunneling at total charge neutrality in
			double-bilayer graphene-{${\mathrm{WSe}}_{2}$} heterostructures},}\ }\href
{\doibase 10.1103/PhysRevLett.120.177702} {\bibfield  {journal} {\bibinfo
		{journal} {Phys. Rev. Lett.}\ }\textbf {\bibinfo {volume} {120}},\ \bibinfo
	{pages} {177702} (\bibinfo {year} {2018})}\BibitemShut {NoStop}%
\bibitem [{\citenamefont {Efimkin}\ \emph {et~al.}(2019)\citenamefont
	{Efimkin}, \citenamefont {Burg}, \citenamefont {Tutuc},\ and\ \citenamefont
	{MacDonald}}]{Efimkin2019}%
\BibitemOpen
\bibfield  {author} {\bibinfo {author} {\bibfnamefont {D.~K.}\ \bibnamefont
		{Efimkin}}, \bibinfo {author} {\bibfnamefont {G.~W.}\ \bibnamefont {Burg}},
	\bibinfo {author} {\bibfnamefont {E.}~\bibnamefont {Tutuc}}, \ and\ \bibinfo
	{author} {\bibfnamefont {A.~H.}\ \bibnamefont {MacDonald}},\ }\bibfield
{title} {\enquote {\bibinfo {title} {Tunneling and fluctuating electron-hole
			{C}ooper pairs in double bilayer graphene},}\ }\href
{https://arxiv.org/abs/1903.07739} {\bibfield  {journal} {\bibinfo  {journal}
		{arXiv:1903.07739v1 [cond-mat.mes-hall]}\ } (\bibinfo {year}
	{2019})}\BibitemShut {NoStop}%
\bibitem [{\citenamefont {Wang}\ \emph {et~al.}(2019)\citenamefont {Wang},
	\citenamefont {Rhodes}, \citenamefont {Watanabe}, \citenamefont {Taniguchi},
	\citenamefont {Hone}, \citenamefont {Shan},\ and\ \citenamefont
	{Mak}}]{Wang2019}%
\BibitemOpen
\bibfield  {author} {\bibinfo {author} {\bibfnamefont {Z.}~\bibnamefont
		{Wang}}, \bibinfo {author} {\bibfnamefont {D.~A.}\ \bibnamefont {Rhodes}},
	\bibinfo {author} {\bibfnamefont {K.}~\bibnamefont {Watanabe}}, \bibinfo
	{author} {\bibfnamefont {T.}~\bibnamefont {Taniguchi}}, \bibinfo {author}
	{\bibfnamefont {J.~C.}\ \bibnamefont {Hone}}, \bibinfo {author}
	{\bibfnamefont {J.}~\bibnamefont {Shan}}, \ and\ \bibinfo {author}
	{\bibfnamefont {K.~F.}\ \bibnamefont {Mak}},\ }\bibfield  {title} {\enquote
	{\bibinfo {title} {Evidence of high-temperature exciton condensation in
			two-dimensional atomic double layers},}\ }\href {\doibase
	10.1038/s41586-019-1591-7} {\bibfield  {journal} {\bibinfo  {journal}
		{Nature}\ }\textbf {\bibinfo {volume} {574}},\ \bibinfo {pages} {76}
	(\bibinfo {year} {2019})}\BibitemShut {NoStop}%
\bibitem [{\citenamefont {Chaves}\ and\ \citenamefont
	{Neilson}(2019)}]{Chaves2019}%
\BibitemOpen
\bibfield  {author} {\bibinfo {author} {\bibfnamefont {A.}~\bibnamefont
		{Chaves}}\ and\ \bibinfo {author} {\bibfnamefont {D.}~\bibnamefont
		{Neilson}},\ }\bibfield  {title} {\enquote {\bibinfo {title} {Two-dimensional
			semiconductors host high-temperature exotic state},}\ }\href {\doibase
	10.1038/d41586-019-02906-9} {\bibfield  {journal} {\bibinfo  {journal}
		{Nature}\ }\textbf {\bibinfo {volume} {574}},\ \bibinfo {pages} {39}
	(\bibinfo {year} {2019})}\BibitemShut {NoStop}%
\bibitem [{\citenamefont {Spielman}\ \emph {et~al.}(2000)\citenamefont
	{Spielman}, \citenamefont {Eisenstein}, \citenamefont {Pfeiffer},\ and\
	\citenamefont {West}}]{Spielman2000}%
\BibitemOpen
\bibfield  {author} {\bibinfo {author} {\bibfnamefont {I.~B.}\ \bibnamefont
		{Spielman}}, \bibinfo {author} {\bibfnamefont {J.~P.}\ \bibnamefont
		{Eisenstein}}, \bibinfo {author} {\bibfnamefont {L.~N.}\ \bibnamefont
		{Pfeiffer}}, \ and\ \bibinfo {author} {\bibfnamefont {K.~W.}\ \bibnamefont
		{West}},\ }\bibfield  {title} {\enquote {\bibinfo {title} {Resonantly
			enhanced tunneling in a double layer quantum {H}all ferromagnet},}\ }\href
{\doibase 10.1103/PhysRevLett.84.5808} {\bibfield  {journal} {\bibinfo
		{journal} {Phys. Rev. Lett.}\ }\textbf {\bibinfo {volume} {84}},\ \bibinfo
	{pages} {5808} (\bibinfo {year} {2000})}\BibitemShut {NoStop}%
\bibitem [{\citenamefont {Comte}\ and\ \citenamefont
	{Nozi\`{e}res}(1982)}]{Comte1982}%
\BibitemOpen
\bibfield  {author} {\bibinfo {author} {\bibfnamefont {C.}~\bibnamefont
		{Comte}}\ and\ \bibinfo {author} {\bibfnamefont {P.}~\bibnamefont
		{Nozi\`{e}res}},\ }\bibfield  {title} {\enquote {\bibinfo {title} {Exciton
			{B}ose condensation : {T}he ground state of an electron-hole gas - {I}.
			{M}ean field description of a simplified model},}\ }\href {\doibase
	10.1051/jphys:019820043070106900} {\bibfield  {journal} {\bibinfo  {journal}
		{J. Phys. France}\ }\textbf {\bibinfo {volume} {43}},\ \bibinfo {pages}
	{1069} (\bibinfo {year} {1982})}\BibitemShut {NoStop}%
\bibitem [{\citenamefont {Keldysh}\ and\ \citenamefont
	{Kopaev}(1965)}]{Keldysh1965}%
\BibitemOpen
\bibfield  {author} {\bibinfo {author} {\bibfnamefont {L.~V.}\ \bibnamefont
		{Keldysh}}\ and\ \bibinfo {author} {\bibfnamefont {Y.~V.}\ \bibnamefont
		{Kopaev}},\ }\bibfield  {title} {\enquote {\bibinfo {title} {Possible
			instability of semimetallic state toward {C}oulomb interaction},}\
}\href@noop {} {\bibfield  {journal} {\bibinfo  {journal} {Sov. Phys. Solid
		State}\ }\textbf {\bibinfo {volume} {6}},\ \bibinfo {pages} {2219} (\bibinfo
{year} {1965})}\BibitemShut {NoStop}%
\bibitem [{\citenamefont {Keldysh}\ and\ \citenamefont
	{Kopaev}(1968)}]{Keldysh1968}%
\BibitemOpen
\bibfield  {author} {\bibinfo {author} {\bibfnamefont {L.~V.}\ \bibnamefont
		{Keldysh}}\ and\ \bibinfo {author} {\bibfnamefont {Y.~V.}\ \bibnamefont
		{Kopaev}},\ }\bibfield  {title} {\enquote {\bibinfo {title} {Collective
			properties of excitons in semiconductors},}\ }\href
{http://jetp.ac.ru/cgi-bin/dn/e_027_03_0521.pdf} {\bibfield  {journal}
	{\bibinfo  {journal} {Sov. Phys. JETP}\ }\textbf {\bibinfo {volume} {27}},\
	\bibinfo {pages} {521} (\bibinfo {year} {1968})}\BibitemShut {NoStop}%
\bibitem [{\citenamefont {Lozovik}\ and\ \citenamefont
	{Yudson}(1975)}]{Lozovik1975}%
\BibitemOpen
\bibfield  {author} {\bibinfo {author} {\bibfnamefont {Y.~E.}\ \bibnamefont
		{Lozovik}}\ and\ \bibinfo {author} {\bibfnamefont {V.~I.}\ \bibnamefont
		{Yudson}},\ }\bibfield  {title} {\enquote {\bibinfo {title} {Feasibility of
			superfluidity of paired spatially separated electrons and holes},}\ }\href
{http://jetpletters.ac.ru/ps/1530/article_23399.pdf} {\bibfield  {journal}
	{\bibinfo  {journal} {JETP Lett}\ }\textbf {\bibinfo {volume} {22}},\
	\bibinfo {pages} {274} (\bibinfo {year} {1975})},\ \bibinfo {note} {(Pis’ma
	Zh.\ Eksp.\ Teor.\ Fiz.\ {\bf 22}, 556 (1975))}\BibitemShut {NoStop}%
\bibitem [{\citenamefont {Lozovik}\ and\ \citenamefont
	{Yudson}(1976)}]{Lozovik1976}%
\BibitemOpen
\bibfield  {author} {\bibinfo {author} {\bibfnamefont {Y.~E.}\ \bibnamefont
		{Lozovik}}\ and\ \bibinfo {author} {\bibfnamefont {V.~I.}\ \bibnamefont
		{Yudson}},\ }\bibfield  {title} {\enquote {\bibinfo {title} {A new mechanism
			for superconductivity: pairing between spatially separated electrons and
			holes},}\ }\href {http://www.jetp.ac.ru/cgi-bin/dn/e_044_02_0389.pdf}
{\bibfield  {journal} {\bibinfo  {journal} {Zh. Eksp. Teor. Fiz}\ }\textbf
	{\bibinfo {volume} {71}},\ \bibinfo {pages} {738} (\bibinfo {year} {1976})},\
\bibinfo {note} {(Sov.\ Phys.\ JETP {\bf 44}, 389 (1976))}\BibitemShut
{NoStop}%
\bibitem [{\citenamefont {Conti}\ \emph
	{et~al.}(2019{\natexlab{b}})\citenamefont {Conti}, \citenamefont {Perali},
	\citenamefont {Peeters},\ and\ \citenamefont {Neilson}}]{Conti2019}%
\BibitemOpen
\bibfield  {author} {\bibinfo {author} {\bibfnamefont {S.}~\bibnamefont
		{Conti}}, \bibinfo {author} {\bibfnamefont {A.}~\bibnamefont {Perali}},
	\bibinfo {author} {\bibfnamefont {F.~M.}\ \bibnamefont {Peeters}}, \ and\
	\bibinfo {author} {\bibfnamefont {D.}~\bibnamefont {Neilson}},\ }\bibfield
{title} {\enquote {\bibinfo {title} {Multicomponent screening and
			superfluidity in gapped electron-hole double bilayer graphene with realistic
			bands},}\ }\href {\doibase 10.1103/PhysRevB.99.144517} {\bibfield  {journal}
	{\bibinfo  {journal} {Phys. Rev. B}\ }\textbf {\bibinfo {volume} {99}},\
	\bibinfo {pages} {144517} (\bibinfo {year} {2019}{\natexlab{b}})}\BibitemShut
{NoStop}%
\bibitem [{\citenamefont {Saberi-Pouya}\ \emph {et~al.}(2018)\citenamefont
	{Saberi-Pouya}, \citenamefont {Zarenia}, \citenamefont {Perali},
	\citenamefont {Vazifehshenas},\ and\ \citenamefont {Peeters}}]{Saberi2018}%
\BibitemOpen
\bibfield  {author} {\bibinfo {author} {\bibfnamefont {S.}~\bibnamefont
		{Saberi-Pouya}}, \bibinfo {author} {\bibfnamefont {M.}~\bibnamefont
		{Zarenia}}, \bibinfo {author} {\bibfnamefont {A.}~\bibnamefont {Perali}},
	\bibinfo {author} {\bibfnamefont {T.}~\bibnamefont {Vazifehshenas}}, \ and\
	\bibinfo {author} {\bibfnamefont {F.~M.}\ \bibnamefont {Peeters}},\
}\bibfield  {title} {\enquote {\bibinfo {title} {High-temperature
		electron-hole superfluidity with strong anisotropic gaps in double
		phosphorene monolayers},}\ }\href {\doibase 10.1103/PhysRevB.97.174503}
{\bibfield  {journal} {\bibinfo  {journal} {Phys. Rev. B}\ }\textbf {\bibinfo
		{volume} {97}},\ \bibinfo {pages} {174503} (\bibinfo {year}
	{2018})}\BibitemShut {NoStop}%
\bibitem [{\citenamefont {Ravensbergen}\ \emph {et~al.}(2018)\citenamefont
	{Ravensbergen}, \citenamefont {Corre}, \citenamefont {Soave}, \citenamefont
	{Kreyer}, \citenamefont {Kirilov},\ and\ \citenamefont
	{Grimm}}]{Ravensbergen2018}%
\BibitemOpen
\bibfield  {author} {\bibinfo {author} {\bibfnamefont {C.}~\bibnamefont
		{Ravensbergen}}, \bibinfo {author} {\bibfnamefont {V.}~\bibnamefont {Corre}},
	\bibinfo {author} {\bibfnamefont {E.}~\bibnamefont {Soave}}, \bibinfo
	{author} {\bibfnamefont {M.}~\bibnamefont {Kreyer}}, \bibinfo {author}
	{\bibfnamefont {E.}~\bibnamefont {Kirilov}}, \ and\ \bibinfo {author}
	{\bibfnamefont {R.}~\bibnamefont {Grimm}},\ }\bibfield  {title} {\enquote
	{\bibinfo {title} {Production of a degenerate {F}ermi-{F}ermi mixture of
			dysprosium and potassium atoms},}\ }\href
{https://journals.aps.org/pra/abstract/10.1103/PhysRevA.98.063624} {\bibfield
	{journal} {\bibinfo  {journal} {Phys. Rev. A}\ }\textbf {\bibinfo {volume}
		{98}},\ \bibinfo {pages} {063624} (\bibinfo {year} {2018})}\BibitemShut
{NoStop}%
\bibitem [{\citenamefont {Pieri}\ \emph {et~al.}(2007)\citenamefont {Pieri},
	\citenamefont {Neilson},\ and\ \citenamefont {Strinati}}]{Pieri2007}%
\BibitemOpen
\bibfield  {author} {\bibinfo {author} {\bibfnamefont {P.}~\bibnamefont
		{Pieri}}, \bibinfo {author} {\bibfnamefont {D.}~\bibnamefont {Neilson}}, \
	and\ \bibinfo {author} {\bibfnamefont {G.~C.}\ \bibnamefont {Strinati}},\
}\bibfield  {title} {\enquote {\bibinfo {title} {Effects of density imbalance
		on the {BCS-BEC} crossover in semiconductor electron-hole bilayers},}\ }\href
{\doibase 10.1103/PhysRevB.75.113301} {\bibfield  {journal} {\bibinfo
		{journal} {Phys. Rev. B}\ }\textbf {\bibinfo {volume} {75}},\ \bibinfo
	{pages} {113301} (\bibinfo {year} {2007})}\BibitemShut {NoStop}%
\bibitem [{\citenamefont {Kinnunen}\ \emph {et~al.}(2018)\citenamefont
	{Kinnunen}, \citenamefont {Baarsma}, \citenamefont {Martikainen},\ and\
	\citenamefont {Törmä}}]{Kinnunen2018}%
\BibitemOpen
\bibfield  {author} {\bibinfo {author} {\bibfnamefont {J.~J.}\ \bibnamefont
		{Kinnunen}}, \bibinfo {author} {\bibfnamefont {J.~E.}\ \bibnamefont
		{Baarsma}}, \bibinfo {author} {\bibfnamefont {J.-P.}\ \bibnamefont
		{Martikainen}}, \ and\ \bibinfo {author} {\bibfnamefont {P.}~\bibnamefont
		{Törmä}},\ }\bibfield  {title} {\enquote {\bibinfo {title} {The
			{F}ulde{\textendash}{F}errell{\textendash}{L}arkin{\textendash}{O}vchinnikov
			state for ultracold fermions in lattice and harmonic potentials: a review},}\
}\href {\doibase 10.1088/1361-6633/aaa4ad} {\bibfield  {journal} {\bibinfo
	{journal} {Rep. Prog. Phys.}\ }\textbf {\bibinfo {volume} {81}},\ \bibinfo
{pages} {046401} (\bibinfo {year} {2018})}\BibitemShut {NoStop}%
\bibitem [{\citenamefont {Frank}\ \emph {et~al.}(2018)\citenamefont {Frank},
	\citenamefont {Lang},\ and\ \citenamefont {Zwerger}}]{Frank2018}%
\BibitemOpen
\bibfield  {author} {\bibinfo {author} {\bibfnamefont {B.}~\bibnamefont
		{Frank}}, \bibinfo {author} {\bibfnamefont {J.}~\bibnamefont {Lang}}, \ and\
	\bibinfo {author} {\bibfnamefont {W.}~\bibnamefont {Zwerger}},\ }\bibfield
{title} {\enquote {\bibinfo {title} {Universal phase diagram and scaling
			functions of imbalanced {F}ermi gases},}\ }\href {\doibase
	10.1134/S1063776118110031} {\bibfield  {journal} {\bibinfo  {journal} {J.
			Exp. Theor. Phys.}\ }\textbf {\bibinfo {volume} {127}},\ \bibinfo {pages}
	{812} (\bibinfo {year} {2018})}\BibitemShut {NoStop}%
\bibitem [{\citenamefont {Wang}\ \emph {et~al.}(2017)\citenamefont {Wang},
	\citenamefont {Che}, \citenamefont {Zhang},\ and\ \citenamefont
	{Chen}}]{Wang2017a}%
\BibitemOpen
\bibfield  {author} {\bibinfo {author} {\bibfnamefont {J.}~\bibnamefont
		{Wang}}, \bibinfo {author} {\bibfnamefont {Y.}~\bibnamefont {Che}}, \bibinfo
	{author} {\bibfnamefont {L.}~\bibnamefont {Zhang}}, \ and\ \bibinfo {author}
	{\bibfnamefont {Q.}~\bibnamefont {Chen}},\ }\bibfield  {title} {\enquote
	{\bibinfo {title} {Enhancement effect of mass imbalance on
			{F}ulde-{F}errell-{L}arkin-{O}vchinnikov type of pairing in {F}ermi-{F}ermi
			mixtures of ultracold quantum gases},}\ }\href
{https://www.nature.com/articles/srep39783} {\bibfield  {journal} {\bibinfo
		{journal} {Sci. Rep.}\ }\textbf {\bibinfo {volume} {7}},\ \bibinfo {pages}
	{39783} (\bibinfo {year} {2017})}\BibitemShut {NoStop}%
\bibitem [{\citenamefont {Forbes}\ \emph {et~al.}(2005)\citenamefont {Forbes},
	\citenamefont {Gubankova}, \citenamefont {Liu},\ and\ \citenamefont
	{Wilczek}}]{Forbes2005}%
\BibitemOpen
\bibfield  {author} {\bibinfo {author} {\bibfnamefont {M.~M.}\ \bibnamefont
		{Forbes}}, \bibinfo {author} {\bibfnamefont {E.}~\bibnamefont {Gubankova}},
	\bibinfo {author} {\bibfnamefont {W.~V.}\ \bibnamefont {Liu}}, \ and\
	\bibinfo {author} {\bibfnamefont {F.}~\bibnamefont {Wilczek}},\ }\bibfield
{title} {\enquote {\bibinfo {title} {Stability criteria for breached-pair
			superfluidity},}\ }\href
{https://journals.aps.org/prl/abstract/10.1103/PhysRevLett.94.017001}
{\bibfield  {journal} {\bibinfo  {journal} {Phys. Rev. Lett.}\ }\textbf
	{\bibinfo {volume} {94}},\ \bibinfo {pages} {017001} (\bibinfo {year}
	{2005})}\BibitemShut {NoStop}%
\bibitem [{\citenamefont {Baarsma}\ and\ \citenamefont
	{Stoof}(2013)}]{Baarsma2013}%
\BibitemOpen
\bibfield  {author} {\bibinfo {author} {\bibfnamefont {J.~E.}\ \bibnamefont
		{Baarsma}}\ and\ \bibinfo {author} {\bibfnamefont {H.~T.~C.}\ \bibnamefont
		{Stoof}},\ }\bibfield  {title} {\enquote {\bibinfo {title} {Inhomogeneous
			superfluid phases in {$^6$Li-$^{40}$K} mixtures at unitarity},}\ }\href
{https://journals.aps.org/pra/abstract/10.1103/PhysRevA.87.063612} {\bibfield
	{journal} {\bibinfo  {journal} {Phys. Rev. A}\ }\textbf {\bibinfo {volume}
		{87}},\ \bibinfo {pages} {063612} (\bibinfo {year} {2013})}\BibitemShut
{NoStop}%
\bibitem [{\citenamefont {Zhang}\ \emph {et~al.}(2019)\citenamefont {Zhang},
	\citenamefont {Maucher},\ and\ \citenamefont {Pohl}}]{Zhang2019}%
\BibitemOpen
\bibfield  {author} {\bibinfo {author} {\bibfnamefont {Y-C.}\ \bibnamefont
		{Zhang}}, \bibinfo {author} {\bibfnamefont {F.}~\bibnamefont {Maucher}}, \
	and\ \bibinfo {author} {\bibfnamefont {T.}~\bibnamefont {Pohl}},\ }\bibfield
{title} {\enquote {\bibinfo {title} {Supersolidity around a critical point in
			dipolar {B}ose-{E}instein condensates},}\ }\href {\doibase
	10.1103/PhysRevLett.123.015301} {\bibfield  {journal} {\bibinfo  {journal}
		{Phys. Rev. Lett.}\ }\textbf {\bibinfo {volume} {123}},\ \bibinfo {pages}
	{015301} (\bibinfo {year} {2019})}\BibitemShut {NoStop}%
\bibitem [{\citenamefont {Neilson}\ \emph {et~al.}(2014)\citenamefont
	{Neilson}, \citenamefont {Perali},\ and\ \citenamefont
	{Hamilton}}]{Neilson2014}%
\BibitemOpen
\bibfield  {author} {\bibinfo {author} {\bibfnamefont {D.}~\bibnamefont
		{Neilson}}, \bibinfo {author} {\bibfnamefont {A.}~\bibnamefont {Perali}}, \
	and\ \bibinfo {author} {\bibfnamefont {A.~R.}\ \bibnamefont {Hamilton}},\
}\bibfield  {title} {\enquote {\bibinfo {title} {Excitonic superfluidity and
		screening in electron-hole bilayer systems},}\ }\href {\doibase
10.1103/PhysRevB.89.060502} {\bibfield  {journal} {\bibinfo  {journal} {Phys.
		Rev. B}\ }\textbf {\bibinfo {volume} {89}},\ \bibinfo {pages} {060502}
(\bibinfo {year} {2014})}\BibitemShut {NoStop}%
\bibitem [{\citenamefont {L\'opez~R\'{\i}os}\ \emph {et~al.}(2018)\citenamefont
	{L\'opez~R\'{\i}os}, \citenamefont {Perali}, \citenamefont {Needs},\ and\
	\citenamefont {Neilson}}]{LopezRios2018}%
\BibitemOpen
\bibfield  {author} {\bibinfo {author} {\bibfnamefont {P.}~\bibnamefont
		{L\'opez~R\'{\i}os}}, \bibinfo {author} {\bibfnamefont {A.}~\bibnamefont
		{Perali}}, \bibinfo {author} {\bibfnamefont {R.~J.}\ \bibnamefont {Needs}}, \
	and\ \bibinfo {author} {\bibfnamefont {D.}~\bibnamefont {Neilson}},\
}\bibfield  {title} {\enquote {\bibinfo {title} {Evidence from quantum
		{M}onte {C}arlo simulations of large-gap superfluidity and {BCS}-{BEC}
		crossover in double electron-hole layers},}\ }\href {\doibase
10.1103/PhysRevLett.120.177701} {\bibfield  {journal} {\bibinfo  {journal}
	{Phys. Rev. Lett.}\ }\textbf {\bibinfo {volume} {120}},\ \bibinfo {pages}
{177701} (\bibinfo {year} {2018})}\BibitemShut {NoStop}%
\bibitem [{\citenamefont {Butov}\ \emph {et~al.}(2002)\citenamefont {Butov},
	\citenamefont {Gossard},\ and\ \citenamefont {Chemla}}]{Butov2002}%
\BibitemOpen
\bibfield  {author} {\bibinfo {author} {\bibfnamefont {L.~V.}\ \bibnamefont
		{Butov}}, \bibinfo {author} {\bibfnamefont {A.~C.}\ \bibnamefont {Gossard}},
	\ and\ \bibinfo {author} {\bibfnamefont {D.~S.}\ \bibnamefont {Chemla}},\
}\bibfield  {title} {\enquote {\bibinfo {title} {Macroscopically ordered
		state in an exciton system},}\ }\href
{https://www.nature.com/articles/nature00943} {\bibfield  {journal} {\bibinfo
		{journal} {Nature}\ }\textbf {\bibinfo {volume} {418}},\ \bibinfo {pages}
	{751} (\bibinfo {year} {2002})}\BibitemShut {NoStop}%
\bibitem [{\citenamefont {High}\ \emph {et~al.}(2012)\citenamefont {High},
	\citenamefont {Leonard}, \citenamefont {Remeika}, \citenamefont {Butov},
	\citenamefont {Hanson},\ and\ \citenamefont {Gossard}}]{High2012}%
\BibitemOpen
\bibfield  {author} {\bibinfo {author} {\bibfnamefont {A.~A.}\ \bibnamefont
		{High}}, \bibinfo {author} {\bibfnamefont {J.~R.}\ \bibnamefont {Leonard}},
	\bibinfo {author} {\bibfnamefont {M.}~\bibnamefont {Remeika}}, \bibinfo
	{author} {\bibfnamefont {L.~V.}\ \bibnamefont {Butov}}, \bibinfo {author}
	{\bibfnamefont {M.}~\bibnamefont {Hanson}}, \ and\ \bibinfo {author}
	{\bibfnamefont {A.~C.}\ \bibnamefont {Gossard}},\ }\bibfield  {title}
{\enquote {\bibinfo {title} {Condensation of excitons in a trap},}\ }\href
{\doibase 10.1021/nl300983n} {\bibfield  {journal} {\bibinfo  {journal} {Nano
			Lett.}\ }\textbf {\bibinfo {volume} {12}},\ \bibinfo {pages} {2605} (\bibinfo
	{year} {2012})}\BibitemShut {NoStop}%
\bibitem [{\citenamefont {Anankine}\ \emph {et~al.}(2017)\citenamefont
	{Anankine}, \citenamefont {Beian}, \citenamefont {Dang}, \citenamefont
	{Alloing}, \citenamefont {Cambril}, \citenamefont {Merghem}, \citenamefont
	{Carbonell}, \citenamefont {Lema\^{\i}tre},\ and\ \citenamefont
	{Dubin}}]{Anankine2017}%
\BibitemOpen
\bibfield  {author} {\bibinfo {author} {\bibfnamefont {R.}~\bibnamefont
		{Anankine}}, \bibinfo {author} {\bibfnamefont {M.}~\bibnamefont {Beian}},
	\bibinfo {author} {\bibfnamefont {S.}~\bibnamefont {Dang}}, \bibinfo {author}
	{\bibfnamefont {M.}~\bibnamefont {Alloing}}, \bibinfo {author} {\bibfnamefont
		{E.}~\bibnamefont {Cambril}}, \bibinfo {author} {\bibfnamefont
		{K.}~\bibnamefont {Merghem}}, \bibinfo {author} {\bibfnamefont {C.~G.}\
		\bibnamefont {Carbonell}}, \bibinfo {author} {\bibfnamefont {A.}~\bibnamefont
		{Lema\^{\i}tre}}, \ and\ \bibinfo {author} {\bibfnamefont {F.}~\bibnamefont
		{Dubin}},\ }\bibfield  {title} {\enquote {\bibinfo {title} {Quantized
			vortices and four-component superfluidity of semiconductor excitons},}\
}\href {\doibase 10.1103/PhysRevLett.118.127402} {\bibfield  {journal}
{\bibinfo  {journal} {Phys. Rev. Lett.}\ }\textbf {\bibinfo {volume} {118}},\
\bibinfo {pages} {127402} (\bibinfo {year} {2017})}\BibitemShut {NoStop}%
\bibitem [{\citenamefont {Timofeev}\ and\ \citenamefont
	{Gorbunov}(2007)}]{Timofeev2007}%
\BibitemOpen
\bibfield  {author} {\bibinfo {author} {\bibfnamefont {V.~B.}\ \bibnamefont
		{Timofeev}}\ and\ \bibinfo {author} {\bibfnamefont {A.~V.}\ \bibnamefont
		{Gorbunov}},\ }\bibfield  {title} {\enquote {\bibinfo {title} {Collective
			state of the {B}ose gas of interacting dipolar excitons},}\ }\href
{https://aip.scitation.org/doi/abs/10.1063/1.2722742} {\bibfield  {journal}
	{\bibinfo  {journal} {J. Appl. Phys.}\ }\textbf {\bibinfo {volume} {101}},\
	\bibinfo {pages} {081708} (\bibinfo {year} {2007})}\BibitemShut {NoStop}%
\bibitem [{\citenamefont {Croxall}\ \emph {et~al.}(2008)\citenamefont
	{Croxall}, \citenamefont {Das~Gupta}, \citenamefont {Nicoll}, \citenamefont
	{Thangaraj}, \citenamefont {Beere}, \citenamefont {Farrer}, \citenamefont
	{Ritchie},\ and\ \citenamefont {Pepper}}]{Croxall2008}%
\BibitemOpen
\bibfield  {author} {\bibinfo {author} {\bibfnamefont {A.~F.}\ \bibnamefont
		{Croxall}}, \bibinfo {author} {\bibfnamefont {K.}~\bibnamefont {Das~Gupta}},
	\bibinfo {author} {\bibfnamefont {C.~A.}\ \bibnamefont {Nicoll}}, \bibinfo
	{author} {\bibfnamefont {M.}~\bibnamefont {Thangaraj}}, \bibinfo {author}
	{\bibfnamefont {H.~E.}\ \bibnamefont {Beere}}, \bibinfo {author}
	{\bibfnamefont {I.}~\bibnamefont {Farrer}}, \bibinfo {author} {\bibfnamefont
		{D.~A.}\ \bibnamefont {Ritchie}}, \ and\ \bibinfo {author} {\bibfnamefont
		{M.}~\bibnamefont {Pepper}},\ }\bibfield  {title} {\enquote {\bibinfo {title}
		{Anomalous {C}oulomb drag in electron-hole bilayers},}\ }\href {\doibase
	10.1103/PhysRevLett.101.246801} {\bibfield  {journal} {\bibinfo  {journal}
		{Phys. Rev. Lett.}\ }\textbf {\bibinfo {volume} {101}},\ \bibinfo {pages}
	{246801} (\bibinfo {year} {2008})}\BibitemShut {NoStop}%
\bibitem [{\citenamefont {Seamons}\ \emph {et~al.}(2009)\citenamefont
	{Seamons}, \citenamefont {Morath}, \citenamefont {Reno},\ and\ \citenamefont
	{Lilly}}]{Seamons2009}%
\BibitemOpen
\bibfield  {author} {\bibinfo {author} {\bibfnamefont {J.~A.}\ \bibnamefont
		{Seamons}}, \bibinfo {author} {\bibfnamefont {C.~P.}\ \bibnamefont {Morath}},
	\bibinfo {author} {\bibfnamefont {J.~L.}\ \bibnamefont {Reno}}, \ and\
	\bibinfo {author} {\bibfnamefont {M.~P.}\ \bibnamefont {Lilly}},\ }\bibfield
{title} {\enquote {\bibinfo {title} {Coulomb drag in the exciton regime in
			electron-hole bilayers},}\ }\href {\doibase 10.1103/PhysRevLett.102.026804}
{\bibfield  {journal} {\bibinfo  {journal} {Phys. Rev. Lett.}\ }\textbf
	{\bibinfo {volume} {102}},\ \bibinfo {pages} {026804} (\bibinfo {year}
	{2009})}\BibitemShut {NoStop}%
\bibitem [{\citenamefont {Vignale}\ and\ \citenamefont
	{MacDonald}(1996)}]{Vignale1996}%
\BibitemOpen
\bibfield  {author} {\bibinfo {author} {\bibfnamefont {G.}~\bibnamefont
		{Vignale}}\ and\ \bibinfo {author} {\bibfnamefont {A.~H.}\ \bibnamefont
		{MacDonald}},\ }\bibfield  {title} {\enquote {\bibinfo {title} {Drag in
			paired electron-hole layers},}\ }\href {\doibase 10.1103/PhysRevLett.76.2786}
{\bibfield  {journal} {\bibinfo  {journal} {Phys. Rev. Lett.}\ }\textbf
	{\bibinfo {volume} {76}},\ \bibinfo {pages} {2786} (\bibinfo {year}
	{1996})}\BibitemShut {NoStop}%
\bibitem [{\citenamefont {Zheng}\ \emph {et~al.}(2016)\citenamefont {Zheng},
	\citenamefont {Croxall}, \citenamefont {Waldie}, \citenamefont {Gupta},
	\citenamefont {Sfigakis}, \citenamefont {Farrer}, \citenamefont {Beere},\
	and\ \citenamefont {Ritchie}}]{Zheng2016}%
\BibitemOpen
\bibfield  {author} {\bibinfo {author} {\bibfnamefont {B.}~\bibnamefont
		{Zheng}}, \bibinfo {author} {\bibfnamefont {A.~F.}\ \bibnamefont {Croxall}},
	\bibinfo {author} {\bibfnamefont {J.}~\bibnamefont {Waldie}}, \bibinfo
	{author} {\bibfnamefont {K.~Das}\ \bibnamefont {Gupta}}, \bibinfo {author}
	{\bibfnamefont {F.}~\bibnamefont {Sfigakis}}, \bibinfo {author}
	{\bibfnamefont {I.}~\bibnamefont {Farrer}}, \bibinfo {author} {\bibfnamefont
		{H.~E.}\ \bibnamefont {Beere}}, \ and\ \bibinfo {author} {\bibfnamefont
		{D.~A.}\ \bibnamefont {Ritchie}},\ }\bibfield  {title} {\enquote {\bibinfo
		{title} {Switching between attractive and repulsive
			{C}oulomb-interaction-mediated drag in an ambipolar {GaAs}/{AlGaAs} bilayer
			device},}\ }\href {\doibase 10.1063/1.4941760} {\bibfield  {journal}
	{\bibinfo  {journal} {Appl. Phys. Lett.}\ }\textbf {\bibinfo {volume}
		{108}},\ \bibinfo {pages} {062102} (\bibinfo {year} {2016})}\BibitemShut
{NoStop}%
\bibitem [{\citenamefont {Tan}\ \emph {et~al.}(1990)\citenamefont {Tan},
	\citenamefont {Snider}, \citenamefont {Chang},\ and\ \citenamefont
	{Hu}}]{Tan1990}%
\BibitemOpen
\bibfield  {author} {\bibinfo {author} {\bibfnamefont {I.~H.}\ \bibnamefont
		{Tan}}, \bibinfo {author} {\bibfnamefont {G.~L.}\ \bibnamefont {Snider}},
	\bibinfo {author} {\bibfnamefont {L.~D.}\ \bibnamefont {Chang}}, \ and\
	\bibinfo {author} {\bibfnamefont {E.~L.}\ \bibnamefont {Hu}},\ }\bibfield
{title} {\enquote {\bibinfo {title} {A self‐consistent solution of
			{S}chr\"odinger--{P}oisson equations using a nonuniform mesh},}\ }\href
{\doibase 10.1063/1.346245} {\bibfield  {journal} {\bibinfo  {journal} {J.
			Appl. Phys.}\ }\textbf {\bibinfo {volume} {68}},\ \bibinfo {pages} {4071}
	(\bibinfo {year} {1990})}\BibitemShut {NoStop}%
\bibitem [{\citenamefont {Rashba}(1960)}]{Rashba1960}%
\BibitemOpen
\bibfield  {author} {\bibinfo {author} {\bibfnamefont {E.~I}\ \bibnamefont
		{Rashba}},\ }\bibfield  {title} {\enquote {\bibinfo {title} {Properties of
			semiconductors with an extremum loop. {I}. {C}yclotron and combinational
			resonance in a magnetic field perpendicular to the plane of the loop},}\
}\href@noop {} {\bibfield  {journal} {\bibinfo  {journal} {Sov. Phys., Solid
		State}\ }\textbf {\bibinfo {volume} {2}},\ \bibinfo {pages} {1109} (\bibinfo
{year} {1960})}\BibitemShut {NoStop}%
\bibitem [{Sup()}]{SupplementalGaAs1}%
\BibitemOpen
\href@noop {} {}\bibinfo {note} {{S}ee {S}upplementary {M}aterial}\BibitemShut {NoStop}%
\bibitem [{\citenamefont {Kosterlitz}\ and\ \citenamefont
	{Thouless}(1973)}]{Kosterlitz1973}%
\BibitemOpen
\bibfield  {author} {\bibinfo {author} {\bibfnamefont {J.~M.}\ \bibnamefont
		{Kosterlitz}}\ and\ \bibinfo {author} {\bibfnamefont {D.~J.}\ \bibnamefont
		{Thouless}},\ }\bibfield  {title} {\enquote {\bibinfo {title} {Ordering,
			metastability and phase transitions in two-dimensional systems},}\ }\href
{\doibase 10.1088/0022-3719/6/7/010} {\bibfield  {journal} {\bibinfo
		{journal} {J. Phys. C: Solid State}\ }\textbf {\bibinfo {volume} {6}},\
	\bibinfo {pages} {1181} (\bibinfo {year} {1973})}\BibitemShut {NoStop}%
\bibitem [{\citenamefont {Botelho}\ and\ \citenamefont {S\'a~de
		Melo}(2006)}]{Botelho2006}%
\BibitemOpen
\bibfield  {author} {\bibinfo {author} {\bibfnamefont {S.~S.}\ \bibnamefont
		{Botelho}}\ and\ \bibinfo {author} {\bibfnamefont {C.~A.~R.}\ \bibnamefont
		{S\'a~de Melo}},\ }\bibfield  {title} {\enquote {\bibinfo {title}
		{Vortex-antivortex lattice in ultracold fermionic gases},}\ }\href {\doibase
	10.1103/PhysRevLett.96.040404} {\bibfield  {journal} {\bibinfo  {journal}
		{Phys. Rev. Lett.}\ }\textbf {\bibinfo {volume} {96}},\ \bibinfo {pages}
	{040404} (\bibinfo {year} {2006})}\BibitemShut {NoStop}%
\bibitem [{\citenamefont {Lozovik}\ \emph {et~al.}(2012)\citenamefont
	{Lozovik}, \citenamefont {Ogarkov},\ and\ \citenamefont
	{Sokolik}}]{Lozovik2012}%
\BibitemOpen
\bibfield  {author} {\bibinfo {author} {\bibfnamefont {Y.~E.}\ \bibnamefont
		{Lozovik}}, \bibinfo {author} {\bibfnamefont {S.~L.}\ \bibnamefont
		{Ogarkov}}, \ and\ \bibinfo {author} {\bibfnamefont {A.~A.}\ \bibnamefont
		{Sokolik}},\ }\bibfield  {title} {\enquote {\bibinfo {title} {Condensation of
			electron-hole pairs in a two-layer graphene system: {C}orrelation effects},}\
}\href {\doibase 10.1103/PhysRevB.86.045429} {\bibfield  {journal} {\bibinfo
	{journal} {Phys. Rev. B}\ }\textbf {\bibinfo {volume} {86}},\ \bibinfo
{pages} {045429} (\bibinfo {year} {2012})}\BibitemShut {NoStop}%
\bibitem [{\citenamefont {Du}\ \emph {et~al.}(2017)\citenamefont {Du},
	\citenamefont {Li}, \citenamefont {Lou}, \citenamefont {Sullivan},
	\citenamefont {Chang}, \citenamefont {Kono},\ and\ \citenamefont
	{Du}}]{Du2017}%
\BibitemOpen
\bibfield  {author} {\bibinfo {author} {\bibfnamefont {L.}~\bibnamefont
		{Du}}, \bibinfo {author} {\bibfnamefont {X.}~\bibnamefont {Li}}, \bibinfo
	{author} {\bibfnamefont {W.}~\bibnamefont {Lou}}, \bibinfo {author}
	{\bibfnamefont {G.}~\bibnamefont {Sullivan}}, \bibinfo {author}
	{\bibfnamefont {K.}~\bibnamefont {Chang}}, \bibinfo {author} {\bibfnamefont
		{J.}~\bibnamefont {Kono}}, \ and\ \bibinfo {author} {\bibfnamefont {R-R.}\
		\bibnamefont {Du}},\ }\bibfield  {title} {\enquote {\bibinfo {title}
		{Evidence for a topological excitonic insulator in {InAs/GaSb} bilayers},}\
}\href {\doibase 10.1038/s41467-017-01988-1} {\bibfield  {journal} {\bibinfo
	{journal} {Nat. Commun.}\ }\textbf {\bibinfo {volume} {8}},\ \bibinfo {pages}
{1971} (\bibinfo {year} {2017})}\BibitemShut {NoStop}%
\bibitem [{\citenamefont {Yang}\ \emph {et~al.}(1997)\citenamefont {Yang},
	\citenamefont {Yang}, \citenamefont {Bennett},\ and\ \citenamefont
	{Shanabrook}}]{Yang1997}%
\BibitemOpen
\bibfield  {author} {\bibinfo {author} {\bibfnamefont {M.~J.}\ \bibnamefont
		{Yang}}, \bibinfo {author} {\bibfnamefont {C.~H.}\ \bibnamefont {Yang}},
	\bibinfo {author} {\bibfnamefont {B.~R.}\ \bibnamefont {Bennett}}, \ and\
	\bibinfo {author} {\bibfnamefont {B.~V.}\ \bibnamefont {Shanabrook}},\
}\bibfield  {title} {\enquote {\bibinfo {title} {Evidence of a hybridization
		gap in ``semimetallic'' {InAs/GaSb} systems},}\ }\href {\doibase
10.1103/PhysRevLett.78.4613} {\bibfield  {journal} {\bibinfo  {journal}
	{Phys. Rev. Lett.}\ }\textbf {\bibinfo {volume} {78}},\ \bibinfo {pages}
{4613} (\bibinfo {year} {1997})}\BibitemShut {NoStop}%
\bibitem [{\citenamefont {Sivan}\ \emph {et~al.}(1992)\citenamefont {Sivan},
	\citenamefont {Solomon},\ and\ \citenamefont {Shtrikman}}]{Sivan1992}%
\BibitemOpen
\bibfield  {author} {\bibinfo {author} {\bibfnamefont {U.}~\bibnamefont
		{Sivan}}, \bibinfo {author} {\bibfnamefont {P.~M.}\ \bibnamefont {Solomon}},
	\ and\ \bibinfo {author} {\bibfnamefont {H.}~\bibnamefont {Shtrikman}},\
}\bibfield  {title} {\enquote {\bibinfo {title} {Coupled electron-hole
		transport},}\ }\href {\doibase 10.1103/PhysRevLett.68.1196} {\bibfield
{journal} {\bibinfo  {journal} {Phys. Rev. Lett.}\ }\textbf {\bibinfo
	{volume} {68}},\ \bibinfo {pages} {1196} (\bibinfo {year}
{1992})}\BibitemShut {NoStop}%
\bibitem [{\citenamefont {Flensberg}\ and\ \citenamefont
	{Hu}(1995)}]{Flensberg1995}%
\BibitemOpen
\bibfield  {author} {\bibinfo {author} {\bibfnamefont {K.}~\bibnamefont
		{Flensberg}}\ and\ \bibinfo {author} {\bibfnamefont {Ben Y.-K.}\ \bibnamefont
		{Hu}},\ }\bibfield  {title} {\enquote {\bibinfo {title} {Plasmon enhancement
			of {C}oulomb drag in double-quantum-well systems},}\ }\href {\doibase
	10.1103/PhysRevB.52.14796} {\bibfield  {journal} {\bibinfo  {journal} {Phys.
			Rev. B}\ }\textbf {\bibinfo {volume} {52}},\ \bibinfo {pages} {14796}
	(\bibinfo {year} {1995})}\BibitemShut {NoStop}%
\bibitem [{\citenamefont {Iskin}\ and\ \citenamefont {S\'a~de
		Melo}(2007)}]{Iskin2007}%
\BibitemOpen
\bibfield  {author} {\bibinfo {author} {\bibfnamefont {M.}~\bibnamefont
		{Iskin}}\ and\ \bibinfo {author} {\bibfnamefont {C.~A.~R.}\ \bibnamefont
		{S\'a~de Melo}},\ }\bibfield  {title} {\enquote {\bibinfo {title} {Mixtures
			of ultracold fermions with unequal masses},}\ }\href {\doibase
	10.1103/PhysRevA.76.013601} {\bibfield  {journal} {\bibinfo  {journal} {Phys.
			Rev. A}\ }\textbf {\bibinfo {volume} {76}},\ \bibinfo {pages} {013601}
	(\bibinfo {year} {2007})}\BibitemShut {NoStop}%
\bibitem [{\citenamefont {Sodemann}\ \emph {et~al.}(2012)\citenamefont
	{Sodemann}, \citenamefont {Pesin},\ and\ \citenamefont
	{MacDonald}}]{Sodemann2012}%
\BibitemOpen
\bibfield  {author} {\bibinfo {author} {\bibfnamefont {I.}~\bibnamefont
		{Sodemann}}, \bibinfo {author} {\bibfnamefont {D.~A.}\ \bibnamefont {Pesin}},
	\ and\ \bibinfo {author} {\bibfnamefont {A.~H.}\ \bibnamefont {MacDonald}},\
}\bibfield  {title} {\enquote {\bibinfo {title} {Interaction-enhanced
		coherence between two-dimensional {D}irac layers},}\ }\href {\doibase
10.1103/PhysRevB.85.195136} {\bibfield  {journal} {\bibinfo  {journal} {Phys.
		Rev. B}\ }\textbf {\bibinfo {volume} {85}},\ \bibinfo {pages} {195136}
(\bibinfo {year} {2012})}\BibitemShut {NoStop}%
\bibitem [{\citenamefont {Gortel}\ and\ \citenamefont {\ifmmode~\acute{S}\else
		\'{S}\fi{}wierkowski}(1996)}]{Gortel1996}%
\BibitemOpen
\bibfield  {author} {\bibinfo {author} {\bibfnamefont {Z.~W.}\ \bibnamefont
		{Gortel}}\ and\ \bibinfo {author} {\bibfnamefont {L.}~\bibnamefont
		{\ifmmode~\acute{S}\else \'{S}\fi{}wierkowski}},\ }\bibfield  {title}
{\enquote {\bibinfo {title} {Superfluid ground state in electron-hole double
			layer systems},}\ }\href {\doibase 10.1016/0039-6028(96)00355-X} {\bibfield
	{journal} {\bibinfo  {journal} {Surf. Sci.}\ }\textbf {\bibinfo {volume}
		{361}},\ \bibinfo {pages} {146} (\bibinfo {year} {1996})}\BibitemShut
{NoStop}%
\bibitem [{\citenamefont {Bistritzer}\ \emph {et~al.}(2008)\citenamefont
	{Bistritzer}, \citenamefont {Min}, \citenamefont {Su},\ and\ \citenamefont
	{MacDonald}}]{Bistritzer2008}%
\BibitemOpen
\bibfield  {author} {\bibinfo {author} {\bibfnamefont {R.}~\bibnamefont
		{Bistritzer}}, \bibinfo {author} {\bibfnamefont {H.}~\bibnamefont {Min}},
	\bibinfo {author} {\bibfnamefont {J.-J.}\ \bibnamefont {Su}}, \ and\ \bibinfo
	{author} {\bibfnamefont {A.~H.}\ \bibnamefont {MacDonald}},\ }\bibfield
{title} {\enquote {\bibinfo {title} {Comment on "{E}lectron screening and
			excitonic condensation in double-layer graphene systems”},}\ }\href
{https://arxiv.org/abs/0810.0331} {\bibfield  {journal} {\bibinfo  {journal}
		{ArXiv:cond-mat/0810.0331v1 [cond-mat.mes-hall]}\ } (\bibinfo {year}
	{2008})}\BibitemShut {NoStop}%
\bibitem [{\citenamefont {Salasnich}\ \emph {et~al.}(2005)\citenamefont
	{Salasnich}, \citenamefont {Manini},\ and\ \citenamefont
	{Parola}}]{Salasnich2005}%
\BibitemOpen
\bibfield  {author} {\bibinfo {author} {\bibfnamefont {L.}~\bibnamefont
		{Salasnich}}, \bibinfo {author} {\bibfnamefont {N.}~\bibnamefont {Manini}}, \
	and\ \bibinfo {author} {\bibfnamefont {A.}~\bibnamefont {Parola}},\
}\bibfield  {title} {\enquote {\bibinfo {title} {Condensate fraction of a
		{F}ermi gas in the {BCS-BEC} crossover},}\ }\href {\doibase
10.1103/PhysRevA.72.023621} {\bibfield  {journal} {\bibinfo  {journal} {Phys.
		Rev. A}\ }\textbf {\bibinfo {volume} {72}},\ \bibinfo {pages} {023621}
(\bibinfo {year} {2005})}\BibitemShut {NoStop}%
\bibitem [{\citenamefont {Guidini}\ and\ \citenamefont
	{Perali}(2014)}]{Guidini2014}%
\BibitemOpen
\bibfield  {author} {\bibinfo {author} {\bibfnamefont {A.}~\bibnamefont
		{Guidini}}\ and\ \bibinfo {author} {\bibfnamefont {A.}~\bibnamefont
		{Perali}},\ }\bibfield  {title} {\enquote {\bibinfo {title} {Band-edge
			{BCS}{\textendash}{BEC} crossover in a two-band superconductor: physical
			properties and detection parameters},}\ }\href
{https://iopscience.iop.org/article/10.1088/0953-2048/27/12/124002/meta}
{\bibfield  {journal} {\bibinfo  {journal} {Supercond. Sci. Tech.}\ }\textbf
	{\bibinfo {volume} {27}},\ \bibinfo {pages} {124002} (\bibinfo {year}
	{2014})}\BibitemShut {NoStop}%
\bibitem [{\citenamefont {Pistolesi}\ and\ \citenamefont
	{Strinati}(1996)}]{Pistolesi1996}%
\BibitemOpen
\bibfield  {author} {\bibinfo {author} {\bibfnamefont {F.}~\bibnamefont
		{Pistolesi}}\ and\ \bibinfo {author} {\bibfnamefont {G.~C.}\ \bibnamefont
		{Strinati}},\ }\bibfield  {title} {\enquote {\bibinfo {title} {Evolution from
			{BCS} superconductivity to {B}ose condensation: {C}alculation of the
			zero-temperature phase coherence length},}\ }\href {\doibase
	10.1103/PhysRevB.53.15168} {\bibfield  {journal} {\bibinfo  {journal} {Phys.
			Rev. B}\ }\textbf {\bibinfo {volume} {53}},\ \bibinfo {pages} {15168}
	(\bibinfo {year} {1996})}\BibitemShut {NoStop}%
\end{thebibliography}
%

\end{document}